\documentclass[1p,twocolumn,12pt]{elsarticle}

\usepackage{graphicx}
\usepackage{amsmath}
\usepackage{amssymb}
\usepackage{bm}
\usepackage{lineno}
\usepackage{url}
\usepackage{color}
\usepackage{soul}

\definecolor{bur}{rgb}{0.8, 0.0, 0.0}

\newcommand\beq{\begin{equation}}
\newcommand\eeq{\end{equation}}
\newcommand\beqa{\begin{eqnarray}}
\newcommand\eeqa{\end{eqnarray}}

\begin{document}

\begin{frontmatter}
%% Title
\title{Benchmarking of a preliminary MFiX-Exa code}

%% authors 
\author[NETL,LRST]{William D. Fullmer}
\author[LBL]{Ann S. Almgren}
\author[LBL]{Michele Rosso}
\author[LBL]{Johannes Blaschke}
\author[NETL]{Jordan Musser\thanks{Corresponding author.}}\ead{jordan.musser@netl.doe.gov}
%% & affiliations
\address[NETL]{National Energy Technology Laboratory, Morgantown, WV 26507, USA}
\address[LRST]{Leidos Research Support Team, Morgantown, WV 26507, USA}
\address[LBL]{Lawrence Berkeley National Laboratory, Berkeley, CA 94720, USA}

%abstract 
\begin{abstract}
MFiX-Exa is a new code being actively developed at Lawrence Berkeley National Laboratory and the National Energy Technology Laboratory as part of the U.S. Department of Energy's Exascale Computing Project. The starting point for the MFiX-Exa code development was the extraction of basic computational fluid dynamic (CFD) and discrete element method (DEM) capabilities from the existing MFiX-DEM code which was refactored into an AMReX code architecture, herein referred to as the preliminary MFiX-Exa code. Although drastic changes to the codebase will be required to produce an exascale capable application, benchmarking of the originating code helps to establish a valid start point for future development. In this work, four benchmark cases are considered, each corresponding to experimental data sets with history of CFD-DEM validation. We find that the preliminary MFiX-Exa code compares favorably with classic MFiX-DEM simulation predictions for three slugging/bubbling fluidized beds and one spout-fluid bed. Comparison to experimental data is also acceptable (within accuracy expected from previous CFD-DEM benchmarking and validation exercises) which is comprised of several measurement techniques including particle tracking velocimetry, positron emission particle tracking and magnetic resonance imaging. The work concludes with an overview of planned developmental work and potential benchmark cases to validate new MFiX-Exa capabilities. 
\end{abstract}
%keywords
\end{frontmatter}
%draft
%\linenumbers

\section{Introduction}
\label{sec.intro}
Owing to its reduced closure modeling and rich, high-fidelity data, coupled computational fluid dynamics and discrete element method (CFD-DEM) has become one of the most commonly applied numerical methods for the simulation of particle fluidization and related gas-solids multiphase flows \cite{zhu08}. There has been a push by the U.S. Department of Energy's (DOE) National Energy Technology Laboratory (NETL) and other institutions to bring this promising technology to industrially relevant problems \cite{cocco17}. Unfortunately, the benefits of CFD-DEM come with associated drawbacks. Because the motion and collision of all particles in a given system is resolved explicitly, CFD-DEM is a computationally expensive technique which only increases with system size, making CFD-DEM simulations at pilot and industrial scales a seriously challenging problem.

Recently, the challenge of CFD-DEM at scale received a significant boost with the selection of MFiX-Exa as one of the US DOE's Exascale Computing Projects (\url{www.exascaleproject.org}). MFiX-Exa aims to drastically reformulate the existing open source MFiX-DEM (\url{mfix.netl.doe.gov}) to deliver an exascale capable CFD-DEM simulation application by 2023. The first step in the construction of MFiX-Exa was to extract only the numerical models required for cold-flow CFD-DEM from MFiX, a general purpose multiphase flow CFD code. These code segments were then refactored into a preliminary MFiX-Exa code using the AMReX framework. AMReX (\url{amrex-codes.github.io}) is a publicly available software framework designed for building massively parallel block structured adaptive mesh refinement applications.

While the focus of MFiX-Exa remains ongoing code development, this work seeks to benchmark the preliminary MFiX-Exa code, i.e., the original code base extracted from MFiX and refactored into the AMReX framework. Although the eventual exascale capable code will appear significantly different, benchmarking of the preliminary MFiX-Exa code helps to establish a validated starting point for ongoing and future code development. The remainder of this manuscript is summarized as follows. In Sec.~\ref{sec.model} the basic governing equations of the preliminary MFiX-Exa code are reviewed. The numerical method is discussed in Sec.~\ref{sec.num} along with general modeling strategies which are employed. The results of four benchmark tests are provided and discussed in Sec.~\ref{sec.results}. Finally, the work closes with a brief review in Sec.~\ref{sec.summary} and an outlook to future MFiX-Exa benchmarking activities in Sec.~\ref{sec.todo}.

\section{Model}
\label{sec.model}
CFD-DEM is an Euler-Lagrange numerical method for multiphase flows in which the dispersed phase, here solid particles, are not resolved by the CFD-grid \cite{tenneti14}. Instead, the dispersed-continuous interaction is modeled via interfacial transfer laws. The motion (and collisions) of all particles are typically solved by either hard-sphere methods (event driven, instantaneous contact), soft-sphere methods (time marching, enduring collisions) \cite{vanderhoef08,deen07}, or some hybrid of the two \cite{lu17b}. MFiX uses a soft-sphere approach which is better suited for enduring and multi-particle contacts often encountered in the dense regions of fluidized beds. For completeness, the governing equations are provided in this section. However, since the preliminary MFiX-Exa code is largely a refactoring of the MFiX DEM code, readers are also referred to the documentation of the original model \citet{garg12, garg12b}. Minor differences between the current implementation and the original MFiX code will be highlighted.

\subsection{CFD}
\label{sec.model.cfd}
The isothermal gas-phase governing equations \citep{garg12} for mass and momentum conservation in the absence of phase change are 
\beq
\frac{ \partial \varepsilon_g \rho_g}{\partial t} + \nabla  \cdot \varepsilon_g \boldsymbol{U}_g = 0,
\label{eq.cont.g}
\eeq
and 
\beq
\frac{ \partial \varepsilon_g \rho_g \boldsymbol{U}_g}{\partial t} 
+ \nabla  \cdot \varepsilon_g \rho_g \boldsymbol{U}_g \otimes \boldsymbol{U}_g
= - \nabla {p_g} + \nabla  \cdot \boldsymbol{\sigma }_g
+ \boldsymbol{M}_{sg} + {\rho _g}{\varepsilon_g}{\boldsymbol{g}},
\label{eq.mom.g}
\eeq
respectively, where $\varepsilon_g$, $\rho_g$, $\boldsymbol{U}_g$, $p_g$ and ${\boldsymbol \sigma }_g$ are the gas-phase volume fraction, material gas density, velocity vector, pressure, and stress tensor, respectively. Modeling details of the generalized interfacial momentum transfer from the solids-phase to the gas-phase, $\boldsymbol{M}_{sg}$, are reserved for Sec.~\ref{sec.model.couple}. The only body force considered is due to gravity, $\boldsymbol{g}$. The gas-phase viscous stress tensor is taken as 
\beq
\boldsymbol{\sigma}_g = \mu_{eff} \left[ \nabla \boldsymbol{U}_g + \left( \nabla \boldsymbol{U}_g \right)^\intercal \right] + \lambda_{eff} \left( \nabla \cdot \boldsymbol{U}_g \right) \boldsymbol{I} 
\label{eq.stress}
\eeq
where $\mu_{eff}$ is an effective dynamic viscosity, $\lambda_{eff}$ is an effective bulk viscosity assumed to be $\lambda_{eff} = -2 \mu_{eff} / 3$ and $\boldsymbol{I}$ is the identity matrix. Currently, stresses due to particle \cite{capecelatro13} and/or shear \cite{vreman09} induced turbulence are neglected such that $\mu_{eff} = \mu_g$, the thermodynamic viscosity.

\subsection{DEM}
\label{sec.model.dem}
The DEM model for the particulate phase solves Newton's laws of motion for each particle. Considering an individual particle, $i$, we have 
\beq
\frac{d \boldsymbol{X}_i}{dt} = \boldsymbol{V}_i ,
\label{eq.xp}
\eeq
\beq
m_i \frac{d \boldsymbol{V}_i}{dt} = m_i \boldsymbol{g} + \boldsymbol{F}_{gi} + \sum_{j=1}^{N_i^{(c)}} \boldsymbol{F}_{ji} ,
\label{eq.vp}
\eeq
and
\beq
I_i \frac{d \boldsymbol{\omega}_i}{dt} = \sum_{j=1}^{N_i^{(c)}} \boldsymbol{T}_{ji} , 
\label{eq.omegap}
\eeq
where $m_i$, $\boldsymbol{X}_i$, $\boldsymbol{V}_i$, $I_i$, and $\boldsymbol{\omega}_i$ are the mass, position, translational velocity, moment of inertia, and angular velocity of the $i^\textrm{th}$ particle, respectively. All particles are assumed to be spherical so that $m_i \equiv \pi \rho_i d_i^3 / 6$ and $I_i = m_i d_i^2 / 10$ where $\rho_i$ and $d_i$ are the density and diameter of the $i^\textrm{th}$ particle, respectively. This work is restricted to monodisperse cases where all $i \in \left[ 1, N \right]$ particles have the same size, $d_i = d_p$, and density $\rho_i = \rho_p$, and, hence, the same mass, $m_i = m$ and moment of inertia, $I_i = I$. The contact force and torque between the $j^\textrm{th}$ and $i^\textrm{th}$ particle are given by $\boldsymbol{F}_{ji}$ and $\boldsymbol{T}_{ji}$ which are summed over all $N_i^{(c)}$ particles and walls in contact with the $i^\textrm{th}$ particle. The collisions terms are closed with a simple soft-sphere contact model in Sec.~\ref{sec.model.lsd}. Finally, details of the interfacial momentum transfer force from the gas-phase to the $i^\textrm{th}$ particle, $\boldsymbol{F}_{gi}$, are provided in Sec.~\ref{sec.model.couple}.

\subsection{Collision Model}
\label{sec.model.lsd}
There are a variety of soft-sphere collision models available in the literature \cite{stevens05},  linear spring dash-pot (LSD) and Hertzian varieties being the most commonly applied models for fluidization \cite{deen07,zhu07}. Both LSD and Hertzian models are available in the MFiX code. However, only the simpler and more computationally efficient LSD model has been extracted and implemented in MFiX-Exa.

Originally owing to Cundall and Strack \cite{cundall79}, the LSD model assumes the normal force acting on the $i^\textrm{th}$ particle by the $j^\textrm{th}$ particle can be described by a conservative spring and a dissipative dash-pot, 
\beq
\boldsymbol{F}_{ji}^{(n)} = -k \delta \boldsymbol{n}_{ji} - \eta \boldsymbol{V}_{ij}^{(n)} ,
\label{eq.lsd.fn}
\eeq
where $k$ and $\eta$ are the spring stiffness and dashpot coefficients, respectively, 
\beq
\delta = r_i + r_j - \left| \boldsymbol{X}_j - \boldsymbol{X}_i \right|
\label{eq.lsd.overlap}
\eeq
is the maximal overlap which must be positive valued for the particles to be in contact,   
\beq
\boldsymbol{n}_{ji} = \frac{\boldsymbol{X}_j - \boldsymbol{X}_i}{\left| \boldsymbol{X}_j - \boldsymbol{X}_i \right|}
\label{eq.lsd.nji}
\eeq
is the normal unit vector pointing to the $j^\textrm{th}$ particle center from the $i^\textrm{th}$ particle center, and
\beq
\boldsymbol{V}_{ij}^{(n)} = \left[ \left( \boldsymbol{V}_i - \boldsymbol{V}_j \right) \cdot \boldsymbol{n}_{ji} \right] \boldsymbol{n}_{ji} 
\label{eq.lsd.vnij}
\eeq
is the normal velocity of the $i^\textrm{th}$ particle relative to the $j^\textrm{th}$ particle. In Eq.~\ref{eq.lsd.overlap}, $r_i = d_i / 2$ and $r_j = d_j / 2$ are the particle radii. In the tangential direction, the computationally efficient model of Capecelatro and Desjardins \cite{capecelatro13} is used,
\beq
\boldsymbol{F}_{ji}^{(t)} = - \mu_{ij} \left| \boldsymbol{F}_{ji}^{(n)} \right| \boldsymbol{t}_{ij} ,
\label{eq.lsd.ft}
\eeq
where $\boldsymbol{t}_{ij}$ is the tangential unit vector taken as 
\beq
\boldsymbol{t}_{ij} = \boldsymbol{V}_{ij}^{(t)} / \left| \boldsymbol{V}_{ij}^{(t)} \right|, 
\label{eq.lsd.tij}
\eeq
where $\boldsymbol{V}_{ij}^{(t)} = \boldsymbol{V}_{ij} - \boldsymbol{V}_{ij}^{(n)}$ is the tangential relative velocity and 
\beq
\boldsymbol{V}_{ij} = \boldsymbol{V}_i - \boldsymbol{V}_j + \left( \ell_{ji}^{(i)} \boldsymbol{\omega}_i + \ell_{ji}^{(j)}  \boldsymbol{\omega}_j \right) \times \boldsymbol{n}_{ji}
\label{eq.lsd.vij}
\eeq
is the total relative velocity at the point of contact for particle $i$ relative to $j$. The torque acting on acting on the $i^\textrm{th}$ particle by being in contact with the $j^\textrm{th}$ particle is given by 
\beq
\boldsymbol{T}_{ji} = \ell_{ji}^{(i)} \boldsymbol{n}_{ji} \times \boldsymbol{F}_{ji}^{(t)}
\label{eq.lsd.Tji}
\eeq
where 
\beq
\ell_{ji}^{(i)} = \frac{\left| \boldsymbol{X}_j - \boldsymbol{X}_i \right|^2 + r_i^2 - r_j^2}{2 \left| \boldsymbol{X}_j - \boldsymbol{X}_i \right|} ,
\label{eq.lsd.lji}
\eeq
is the distance between the $i^{\textrm{th}}$ particle center and the $i$-$j$ contact plane. Likewise, $\ell_{ji}^{(j)} = 
\left| \boldsymbol{X}_j - \boldsymbol{X}_i \right| - \ell_{ji}^{(i)}$ in Eq.~\ref{eq.lsd.vij} is the distance between the $j^{\textrm{th}}$ particle center and the $i$-$j$ contact plane.

A few comments on the simplified model form of Eq.~\ref{eq.lsd.ft} are needed. First, we note that the simplified form differs from the original MFiX code which considered the full tangential collision model including tangential spring and dashpot coefficients \cite{garg12, garg12b, deen07}. The simplified model used in MFiX-Exa is more computationally efficient, because it does not require integrating (and storing) the tangential displacement for all enduring contact pairs. As a result, the simplified model is less accurate at predicting the tangential restitution coefficient of acute particle collisions, see Fig.~7 of Ref.~\cite{capecelatro13}.

\subsection{Wall interactions}
\label{sec.model.wall}

After checking for particle-particle collisions, potential particle-wall collisions are resolved. Unlike the original MFiX which considered planar walls (for rectangular geometries), wall collisions in MFiX-Exa use the Embedded-Boundary (EB) framework native to AMReX. When simulations are initialized, the computational grid is filled with an \texttt{ebflags} array, which indicates whether cells intersect with a wall. At this point, local wall positions and normals are stored. In this context, the wall is broken up into local EB ``facets'', one per cell. When particles test for collisions, the $3^3$ cells surround (and including) the particle's cell are checked for \texttt{ebflags}. If multiple \texttt{ebflags} are detected, each is checked for collisions (particle overlaps with EB facet in that cell). For each overlapping facet, the particle-wall force $\boldsymbol{F}_{i,w_j}$ for particle $i$ colliding with facet $w_j$ is summed into a total particle-wall force $\boldsymbol{F}_{i, w} = \sum_j \boldsymbol{F}_{i, w_j}$. The force model used for each $\boldsymbol{F}_{i, w_j}$ is the same as the particle-particle force model described in the previous section with some small changes. Unlike particle-particle collisions, for particle-wall collisions, the normal of the EB facet $\boldsymbol{n}_j$ is used to determine the direction of $\boldsymbol{F}_{i, w_j}= f_{i,w_j}(\boldsymbol{X}_j)\boldsymbol{n}_j$ where $\boldsymbol{X}_j$ is the closest position to the particle on the EB facet. We note that if this position is at the corner of two or more EB-facets, then $\boldsymbol{n}_j$ points along the line connecting $\boldsymbol{X}_i$ (the particle position) and $\boldsymbol{X}_j$ (the corner position).

\subsection{Coupling}
\label{sec.model.couple}
Following the original MFiX code, the interfacial forces on the $i^\textrm{th}$ particle from the gas phase are taken as the sum of buoyancy and drag, 
\beq
\boldsymbol{F}_{gi} = - \mathcal{V}_i \nabla p_g - \frac{1}{2} C_D \rho_g \boldsymbol{V}_{ig} \left|\boldsymbol{V}_{ig}\right| A_i^{(proj)} , 
\label{eq.couple.fgi}
\eeq
where $C_D$ is the drag coefficient, $\boldsymbol{V}_{ig} = \boldsymbol{V}_i - \boldsymbol{U}_g ( \boldsymbol{X}_i )$ is the velocity of $i^\textrm{th}$ particle relative to the gas-phase (at the position of the $i^\textrm{th}$ particle) and $A_i^{proj}$ is the projected area of the particle. Again assuming spherical particles $A_i^{proj}$ is simply $\pi r_i^2$. Generally, the gas-particle interaction force of Eq.~(\ref{eq.couple.fgi}) should include the gas-phase viscous stress stress tensor and interfacial forces due to velocity gradients (lift force), rotation (Magnus force), acceleration (virtual mass force), and transient boundary layer development (Basset force), among others \cite{ishiihibiki,drew}. Here, we assume the most important interfacial effects are captured with buoyancy (pressure gradient) and steady drag, a common assumption in high density ratio, high Stokes number gas-solids multiphase flow modeling.

Closure for the drag coefficient typically comes from experimental or direct numerical simulation data, e.g., see Beetstra et al. \cite{beetstra07}. The cases studied in Sec.~\ref{sec.results} consider relatively large, Geldart Group D particles \cite{geldart73}. Therefore, we use the empirical drag law proposed by Gidaspow \cite{ding90}, 
\beq
C_D = \chi C_D^{(Wen-Yu)} + \left(1 - \chi \right)  C_D^{(Ergun)} 
\label{eq.drag.gidaspow}
\eeq 
which combines the Wen-Yu \cite{wen66} relation in dilute regions, 
\beq
C_D^{(Wen-Yu)} = \max \left[\frac{24}{Re_i}\left(1 + 0.15 Re_i^{0.687}\right),\ 0.44 \right] \left( 1 - \varepsilon_g \right)^{-1.65} 
\label{eq.drag.wenyu}
\eeq
with the Ergun equation \cite{ergun52} in dense regions,  
\beq
C_D^{(Ergun)} = \frac{200 \left(1 - \varepsilon_g\right)}{Re_i} + \frac{7}{3} ,  
\label{eq.drag.ergun}
\eeq
using the smooth switch proposed by Lathouwers and Bellan \cite{lathouwers01}, 
\beq
\chi = \frac{\arctan 150 \left( \varepsilon_g - 0.8 \right)}{\pi} + \frac{1}{2}.
\label{eq.drag.blend}
\eeq
In Eqs.~\ref{eq.drag.wenyu} and \ref{eq.drag.ergun},   
\beq
Re_i = \frac{\rho_g \left( 1 - \varepsilon_g \right) d_i \left| \boldsymbol{V}_{ig} \right| }{\mu_g} , 
\label{eq.drag.rep}
\eeq
is the $i^\textrm{th}$ particle Reynolds number.

Specification of the drag law effectively closes the system of equations. However, the transfer of point-wise, Lagrangian particle information to the continuous, Eulerian fluid field remains to be specified. In general, the L-E transfer occurs through volume filtering \cite{capecelatro13}
\beq
(1 - \varepsilon_g) A (\boldsymbol{x},t) \approx \sum_{i=1}^{N_p} A_i(\boldsymbol{X}_i,t) {\mathcal G} (\left| \boldsymbol{x} - \boldsymbol{X}_i \right|) {\mathcal V}_i , 
\label{eq.transfer.general}
\eeq
where $A_i$ is a general particle property and ${\mathcal G}$ is a strictly positive, unit normal filtering kernel. The gas volume fraction and generalized interfacial momentum transfer, $\boldsymbol{M}_{sg}$, are determined from Eq.~(\ref{eq.transfer.general}) by setting $A_i$ to unity and $\boldsymbol{F}_{gi}/{\mathcal V_i}$, repectively. In practice, direct application of Eq.~(\ref{eq.transfer.general}) is computationally expensive. Therefore compact, grid-based kernels are applied in MFiX and MFiX-Exa, as discussed in the following Sec.~\ref{sec.num}, so that only a small subset of particles local to $\boldsymbol{x}$ needed to calculate $A (\boldsymbol{x},t)$.

\section{Numerical Solution}
\subsection{Numerical method}
\label{sec.num}
The preliminary MFiX-exa code uses only uniform, rectangular grids to solve the fluid governing equations in a finite volume formulation in the style of Patankar's method for single phase flow \cite{patankar}. Field variables are stored on a staggered grid with pressure and void fraction stored at cell centers, i.e.,  $p_g^{(i,j,k)}$ and $\varepsilon_g^{(i,j,k)}$, and velocity components are staggered about cell faces, i.e., 
$u_g^{(i+1/2,j,k)}$, $v_g^{(i,j+1/2,k)}$, and $w_g^{(i,j,k+1/2)}$. The superscript $(i,j,k)$ indicates the $\boldsymbol{x}^{(i,j,k)} = \left[(i-1/2) dx,\ (j-1/2) dy,\ (k-1/2) dz \right]^\intercal$ grid position where $ds = L_s / N_s$ is the grid spacing, $L_s$ is the domain length and $N_s$ is the number of CFD grid cells in each $s = x$-, $y$-, $z$-direction. For simplicity, only first-order upwinding is retained for variable extrapolation. The scheme is also temporally (formally) first-order accurate with (iterative) backward Euler time stepping. Pressure-velocity coupling is achieved through a multiphase SIMPLE scheme \cite{syamlal98}. The stabilized bi-conjugate gradient (BiCGStab) method is used to solve the matrix equations without preconditioners.

CFD-DEM coupling is explicit, i.e., information is exchanged at the beginning of a timestep, the CFD solver is advanced one CFD timestep, $dt_{CFD}$, then the particles are advanced to the $n+dt_{CFD}$ time-level using the $n$ time-level exchange data. DEM advancement is first-order forward Euler and is typically sub-cycled, i.e., $dt_{DEM} < dt_{CFD}$. Gas-phase pressure gradient is computed using a central difference about the adjacent cells from the cell in which each particle resides. Volume (for the calculation void fraction) and drag force is deposited onto the fluid grid using the Linear Hat transfer kernel \cite{snider98}. Likewise, fluid velocity is  interpolated to particle positions using tri-linear interpolation.

\subsection{Modeling Strategy}
\label{sec.num.strategy}
In this section, we provide some of the guiding principles used in setting up the benchmark cases. One of the most important parameters in any discretized numerical method is the grid spacing. For the monodispersed particulate flows considered here, it is convenient to write the non-dimensional grid spacing as
\beq
\Delta^* = \sqrt[3]{dx \ dy \ dz} / d_p . 
\label{eq.Dstar}
\eeq
Convergence tests from numerous previous CFD-DEM studies have produced a common heuristic: a grid size of $\Delta^* \approx 2$ is required to provide grid-insensitive solutions \cite{cocco17}. Furthermore, a recent solution verification study of a fixed particle assemblies \cite{fullmer18c} indicate that the CFD discretization error is small for $\Delta^* \le 2$. Where possible, we try to adhere to this criteria.

Another source of numerical uncertainty in (soft-sphere) CFD-DEM simulations is the spring constant. In the absence of cohesion \cite{liu16} or heat transfer \cite{morris16b}, particles are typically made as soft as possible while retaining solution insensitivity. The spring constant is related to the collison time scale by
\beq
\tau_{coll} = \pi \left[ \frac{k}{2 \hat{m}_{ij}} - \left(\frac{\eta}{4 \hat{m}_{ij}}\right)^2 \right]^{-1/2} , 
\label{eq.tcoll}
\eeq
where $\hat{m}_{ij} = 2(m_i^{-1} + m_j^{-1})^{-1}$ is the harmonic mean of the mass of two colliding particles $i$ and $j$. In this work, we set $k$ such that $\tau_{coll}$ is much smaller that typical hydrodynamic time scales. Note that by setting $k$, and assuming the restitution coefficient is a (roughly constant) material property, the dashpot coefficient is set from  Eq.~(\ref{eq.tcoll}) and $\ln e = -\eta \tau_{coll} / 4 \hat{m}_{ij}$.

The DEM timestep is then set such that collisions are resolved by 20 steps, i.e., $dt_{DEM} = \tau_{coll}/20$. Because the preliminary MFiX-Exa code uses explicit coupling, the fluid is prevented from advancing more than 20 $dt_{DEM}$ sub-cycles, or $\max dt_{CFD} = \tau_{coll}$. The initial time step is set to $dt_{CFD} = \tau_{coll}$ and an adaptive timestepping algorithm is used which can reduce the timestep if iterative convergence criteria are not satisfied. Although not rigorously studied, a few tests suggest limiting $dt_{CFD}$ to one $\tau_{coll}$ time scale may be overly conservative, at least for simple flows (monodisperse, non-reacting, etc.). More detailed investigation into the deterioration of solution accuracy with increasing $dt_{CFD}/dt_{DEM}$ would be a welcome addition to the collective CFD-DEM knowledge-base.

Although most of the numerical scheme is only formally first-order accurate, with the application of a fine mesh, a small DEM timestep and a small sub-cycling restriction ($dt_{CFD}/dt_{DEM} \le 20$), we assume that the largest source of numerical error is statistical, i.e., due to finite time-averaged statistics. In this work, we use the method of non-overlapping bins to compute confidence intervals (CIs) on the time-averaged data \cite{syamlal17}. Twelve temporal non-overlapping bins are used in each case. Although the CIs are computed for all statistics for all simulations, they are only plotted in Sec.~\ref{sec.results} for the preliminary MFiX-Exa simulations to keep the figures readable.

\section{Results}
\label{sec.results}
The results of four benchmarking exercises are presented in Sec.~\ref{sec.goldschmidt} - Sec.~\ref{sec.sscp} below. All cases are physical problems with experimental data, therefore, this study could be considered validation. However, we are more concerned here with code-to-code comparisons of physical measures rather than analyzing model form error of this CFD-DEM implementation. In other words, deviation from experimental data is acceptable here as long as the discrepancy is inherent to CFD-DEM (or the experimental data) and not due to the code refactoring, hence the more general benchmarking nature of this work. It may also be noted that the classic MFiX code being compared against, described in Sec.~\ref{sec.classic}, has been validated by Li et al. \cite{li12} and additionally in several separate studies, e.g., see \cite{lamarche15, li16, xu18, bakshi18}.

There are four potential sources for differences between the comparisons which follow. First, the models themselves are slightly diffent, e.g., the simplified tangential LSD force of Eq.~\ref{eq.lsd.ft}. The second source is due to differences in the algoriths, e.g., the wall boundary condition discussed in Sec.~\ref{sec.model.wall}. Another source, which would be easy to overlook, is simply due to the specific code implementation, i.e., call sequence, order of operations, etc. Finally, there may also be coding errors and mistakes, i.e., bugs, introduced during the refactoring. This study is primarily focused on uncovering this fourth source of code-to-code disagreement. It is worth noting that the results presented herein are the final results, that is, after several bugs were discovered, identified and fixed through benchmark testing.

\subsection{MFiX classic simulations}
\label{sec.classic}
Since the preliminary MFiX-Exa code was refactored from classic MFiX-DEM, this is the most important code to compare against. MFiX 2016.1 is the closest released MFiX version to when the refactoring began taking place, which is used for reference simulations herein. Although the preliminary MFiX-Exa code and MFiX Release 2016.1 share the same codebase, some minor implementation differences prevent exact model replication. The classic MFiX codebase considered a full tangential collision force model, rather than the simplified model of Eq.~(\ref{eq.lsd.ft}). Therefore normal and tangential LSD parameters need to be specified; here we assume $k_n = k$, $\eta_n = \eta$, $k_t = 2 k_n / 7$ and $\eta_t = \eta_n / 2$. Additionally, although the \texttt{GARG\_2012} transfer kernel \cite{garg07} is quite similar to the linear hat of MFiX-Exa, \texttt{GARG\_2012} in MFiX 2016.1 requires ``implicit'' coupling, i.e., the drag coefficient and fluid velocity are updated for each sub-cycle. Therefore, we consider two slightly different MFiX 2016.1 models: an implicitly coupled model using the \texttt{GARG\_2012} kernel and an explicitly coupled model using \texttt{SQUARE\_DPVM}, a cubic transfer kernel, with an edge length equal to $1.5 d_p$. The CFD timestep of the explicit \texttt{SQUARE\_DPVM} model limited by $dt_{CFD} \le \tau_{coll}$. The ``implicit'' \texttt{GARG\_2012} model is only limited by $dt_{CFD} \le 0.1 s$, however, it is assumed that the actual timestep is convergence limited below this upper limit. By default, MFiX 2016.1 takes $dt_{DES} = \tau_{coll} / 50$ which is increased to $\tau_{coll}/20$. To match the preliminary MFiX-Exa code, first-order upwinding (FOU) is used for variable extrapolation. However, as these results might also be useful to benchmark future MFiX-Exa codes with increased numerical accuracy, a higher-order SMART flux-limiter scheme \cite{gaskell88, waterson07} is also considered.

\subsection{Goldschmidt fluidized bed}
\label{sec.goldschmidt}
Much of the existing CFD-DEM validation data sets are a result of an extensive campaign by J. A. M. Kuipers and colleagues. One of the earliest experiments from this group is the thin (``pseudo-2D''), fluidized bed of Goldschmidt et al. \cite{goldschmidt04}, referred to hereafter as the Goldschmidt bed. In addition to easily measured material properties, collisional properties of the glass beads needed for discrete particle simulations were measured and reported. The bed dimensions, material properties, and collision properties used in the simulations are listed in Table~\ref{t.goldschmidt}. Because the depth of the bed is only resolved by three CFD grid cells, free-slip wall boundary conditions (BCs) are used for the fluid along the front and back walls. The side walls are treated as no-slip. A uniform gas inflow BC is placed at the bottom inlet and a pressure outflow is used at the top exit. The particle bed contains approximately twenty-five thousand glass beads initialized in a randomly distributed array with a small, random initial velocity. The initial fluid field is at rest. The inlet velocity is linearly increased from zero to $U_{in}$ over a period of one second. Three conditions are studied: $U_{in}/U_{mf} = 1.25$, 1.50 and 2.00, where $U_{mf} = 1.25$~m/s is the \emph{measured} minimum fluidization velocity.

\begin{table}[htb]
  \begin{center}
    \caption{Simulation parameters of the Goldschmidt bed.}
    \label{t.goldschmidt}
    \begin{tabular}{lll}
    \hline
    \bf{Bed properties} &  &  \\
    Width & $L_x$ & 150 (mm) \\
    Height & $L_y$ & 700 (mm) \\
    Depth & $L_z$ & 15 (mm) \\
    Grid & $\Delta^*$ & 2.00 \\
    \bf{Particle properties} &  &  \\
    Number & $N_p$ & 24750 \\
    Diameter & $d_{p}$ & 2.49 (mm) \\
    Density & $\rho_{p}$ & 2526 (kg/m$^3$) \\
    \bf{Collision properties} &  &  \\
    Restitution coeff. & $e_{pp}$, $e_{pw}$ & 0.97, 0.97 \\
    Friction coeff. & $\mu_{pp}$, $\mu_{pw}$ & 0.10, 0.09 \\
    Spring stiffness & $k$ & 2519 (N/m) \\
    \bf{Fluid properties} &  &  \\
    Density & $\rho_g$ & 1.2 (kg/m$^3$) \\
    Viscosity & $\mu_g$ & $1.8 \times 10^{-5}$ (Pa-s) \\
    \hline
    \end{tabular}
  \end{center}
\end{table}

In addition to qualitative snapshots, the Goldschmidt bed expansion dynamics were analyzed through video recordings. The bed was recorded at a frequency of 25 Hz from a period of 5 to 60s. Every particle in the frame was then identified though digital image analysis and the elevation of each is averaged to determine the bed height, $h_{bed}(t)$. The bed height is then time-averaged to determine the mean, $\bar{h}_{bed}$, and standard deviation, $h'_{bed}$.

\begin{table}[htb]
  \begin{center}
    \caption{Mean bed height, $\bar{h}_{bed}$, in the Goldschmidt bed. All measurements reported in (mm).}
    \label{t.goldschmidt.hbar}
    \begin{tabular}{llll}
    \hline
    $U_{in} = $              &  $1.25 U_{mf}$    &   $1.50 U_{mf}$     &   $2.00 U_{mf}$   \\
    experiment               &  92               &   114               &   135             \\
    exa.18.08.simple         &  97.4 $\pm$ 0.2   &   100.5 $\pm$ 0.7   &   133.0 $\pm$ 6.1 \\
    mfix.2016.1.garg.fou     &	93.9 $\pm$ 0.4   &   114.3 $\pm$ 1.0   &   140.5 $\pm$ 2.3 \\
    mfix.2016.1.sqdpvm.fou   &	91.3 $\pm$ 0.2   &   103.5 $\pm$ 0.3   &   124.4 $\pm$ 0.8 \\
    mfix.2016.1.garg.smart   &	92.3 $\pm$ 0.2   &   109.1 $\pm$ 1.1   &   136.5 $\pm$ 2.4 \\
    mfix.2016.1.sqdpvm.smart &	93.1 $\pm$ 0.2   &   106.8 $\pm$ 0.8   &   130.8 $\pm$ 1.3 \\
    \hline
    \end{tabular}
  \end{center}
\end{table}

\begin{table}[htb]
  \begin{center}
    \caption{Fluctuating bed height, $h'_{bed}$, in the Goldschmidt bed. All measurements reported in (mm).}
    \label{t.goldschmidt.hstd}
    \begin{tabular}{llll}
    \hline
    $U_{in} = $              &  $1.25 U_{mf}$     &   $1.50 U_{mf}$     &   $2.00 U_{mf}$   \\
    experiment               &  9.8               &  22.6               &  32.3             \\
    exa.18.08.simple         & 12.79 $\pm$ 0.10   &   7.01 $\pm$ 0.90   &  20.38 $\pm$ 5.09 \\
    mfix.2016.1.garg.fou     &  5.97 $\pm$ 0.46   &  13.18 $\pm$ 1.32   &  21.51 $\pm$ 3.36 \\ 
    mfix.2016.1.sqdpvm.fou   &  1.71 $\pm$ 0.10   &   5.31 $\pm$ 0.32   &   9.51 $\pm$ 1.12 \\
    mfix.2016.1.garg.smart   &  4.20 $\pm$ 0.55   &  10.47 $\pm$ 0.98   &  21.11 $\pm$ 3.43 \\
    mfix.2016.1.sqdpvm.smart &  2.52 $\pm$ 0.16   &   7.31 $\pm$ 0.95   &  13.88 $\pm$ 1.34 \\
    \hline
    \end{tabular}
  \end{center}
\end{table}

In the CFD-DEM simulations, two conditions keep particles from being averaged into $h_{bed}$. First, it was reported that a flange obscures the 13 mm above the inlet; therefore, particles with $y_i < 13$~mm are neglected. The second, and more complicated condition, is that the particles near the front of the bed obscure particles behind them. We account for this limitation in the physical depth of view by neglecting particles with $z_i > 3.75$~mm, i.e., $1.5 d_p$. Time-averaging occurs in 5s non-overlapping bins from 5 to 65s simulation time. The results with 95\% CIs are reported in Tables~\ref{t.goldschmidt.hbar} and \ref{t.goldschmidt.hstd}. Generally, the mean bed height predicted by the preliminary MFiX-Exa code is in good agreement with the four classic MFiX results, the experimental data and the original discrete particle simulation results \cite{goldschmidt04} (not shown). One minor discrepancy is that $\bar{h}_{bed}$ for the $1.25 U_{mf}$ case is larger than the rest. The difference in this case is even more noticeable for the bed fluctuation. Although, $h'_{bed}$ predicted by MFiX-Exa at $1.25 U_{mf}$ happens to be the closest to the experimental data, there are a few issues with this data point: it significantly outlies the other four simulation results, it is almost an order of magnitude larger than the original discrete particle simulation results and it breaks the expected trend of increasing $h'_{bed}$ with increasing $U_{in}$. Analysis of the transient $h_{bed}(t)$ shows that the MFiX-Exa result at $U_{in} = 1.25 U_{mf}$ produces an extremely regular bubbling/slugging pattern which appears more chaotic in the other models. We believe that the regularity of hydrodynamic pattern may be largely attributed to the thinness of the bed and is likely confined to a narrow model input parameter space. It will be interesting to test future MFiX-Exa codes at this condition to see if a higher-order scheme is sufficient to produce a more chaotic bubbling/slugging pattern and lower the fluctuating bed height measurement.

%%
%%  mueller
%%
\subsection{M{\"u}ller fluidized bed}
\label{sec.mueller}
The second validation case reported here is taken from the experiments of M{\"u}ller et al. \cite{mueller08, mueller09}
which have been widely used for validation of CFD-DEM models, including the original classic MFiX-DEM implementation \cite{li12}.
The experiments consist of a thin, ``pseudo-2D'' clear bed filled with poppy seeds. Due to the moisture in the seeds, high-speed spatio-temporal data of the bed concentration and velocity can be extracted using magnetic resonance imaging (MRI). However, it should be noted that several fundamental assumptions of the CFD-DEM model are stressed in this case due to the irregularity of the particles, i.e., the seeds. The bed is fluidized by a uniform inflow at superficial velocities of $U/U_{mf}$ = 2 and 3. Here, we select only the $3U_{mf}$ case as profiles for both void fraction at two elevations \cite{mueller09} and particle velocity at three elevations \cite{mueller08} were reported for this case, which is hereafter referred to as the M{\"u}ller bed.

\begin{table}[htb]
  \begin{center}
    \caption{Simulation parameters of the M{\"u}ller bed.}
    \label{t.mueller}
    \begin{tabular}{lll}
    \hline
      \bf{Bed properties} &  &  \\
      Width & $L_x$ & 44 (mm) \\
      Height & $L_y$ & 120 (mm) \\
      Depth & $L_z$ & 10 (mm) \\
      Grid & $\Delta^*$ & 2.07 \\
      \bf{Particle properties} &  &  \\
      Number & $N_p$ & 9240 \\
      Diameter & $d_{p}$ & 1.2 (mm) \\
      Density & $\rho_{p}$ & 1000 (kg/m$^3$) \\
      \bf{Collision properties} &  &  \\
      Restitution coeff. & $e_{pp}$, $e_{pw}$ & 0.97 \\
      Friction coeff. & $\mu_{pp}$, $\mu_{pw}$ & 0.1 \\
      Spring stiffness & $k_n$ & 440 (N/m) \\
      \bf{Fluid properties} &  &  \\
      Density & $\rho_g$ & 1.2 (kg/m$^3$) \\
      Viscosity & $\mu_g$ & $1.8 \times 10^{-5}$ (Pa-s) \\
      \hline
    \end{tabular}
  \end{center}
\end{table}

The details of the simulation parameters are provided in Table~\ref{t.mueller}. A grid of $N_x \times N_y \times N_z =  18 \times 48 \times 4$ is applied which gives a dimensionless grid spacing of $\Delta^* \approx 2$. Due to the low resolution of the bed depth, free-slip BCs are applied to the front and back walls while no-slip BCs are applied at the right and left walls. Uniform, pure-gas mass inflow of $U_g = 0.9$ (m/s) is applied at the inlet and a constant pressure BC at the outlet.  Void fraction profiles are computed from the CFD-grid causing a slight discrepancy in the location of the data compared to the experiments (18 \emph{vs.} 22 $x$-locations). The particle velocities are bin averaged into the same $x-y$ regions as in the experiment using straightforward centroid deposition (top-hat averaging). Both void fraction and particle velocity profiles are averaged across the bed depth, consistent with the MRI technique. Simulations are run for 65s which take roughly a day of wall clock time in serial. The method of non-overlapping batch means is used to determine an appropriate averaging region of approximately 5s. The first 5s of data is discarded as initial start up transient and the remaining twelve 5s intervals are averaged to determine mean and 95\% CIs. Again, all model CIs are quite similar and only one is reported as a gauge of the statistical (time-averaging) uncertainty present in the numerical results.

\begin{figure}[htb]
\centering\includegraphics[height=0.3\linewidth]{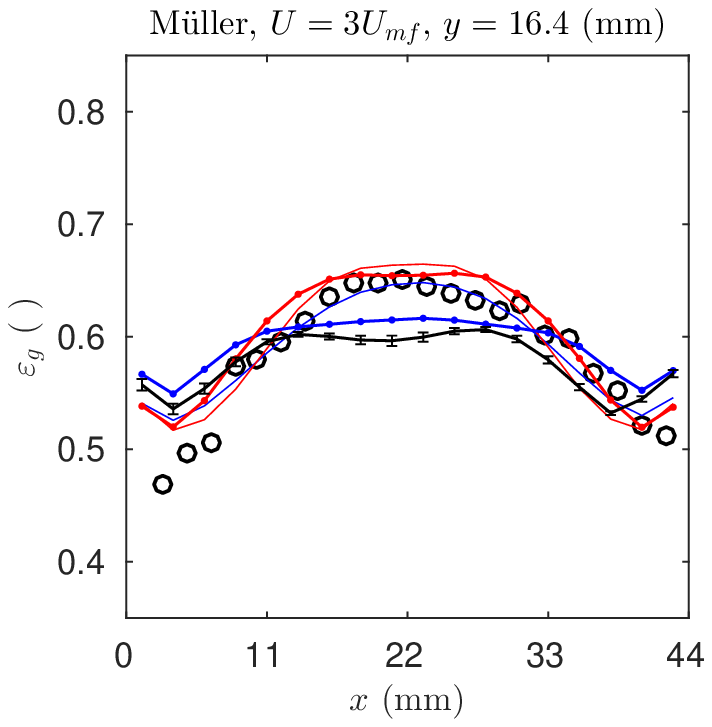}
\centering\includegraphics[height=0.3\linewidth]{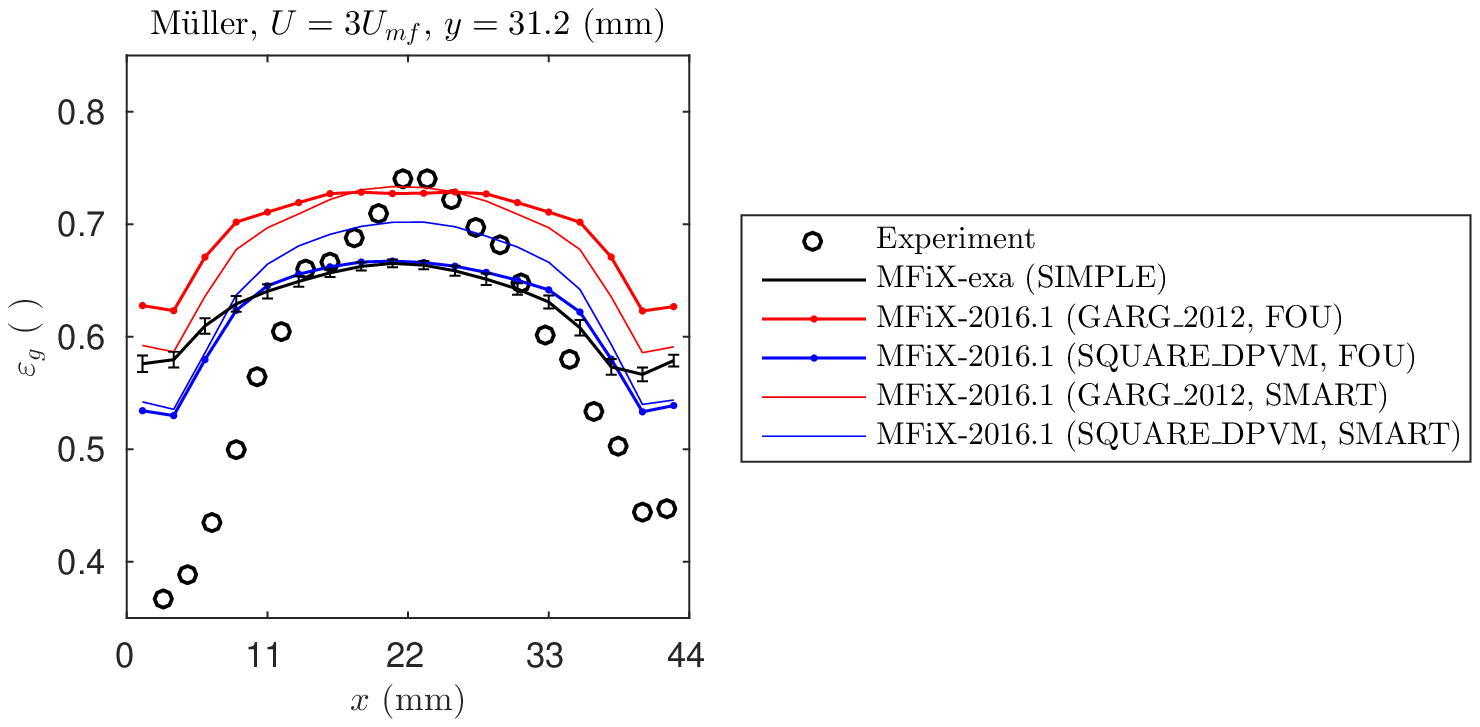} \\
\centering\includegraphics[height=0.3\linewidth]{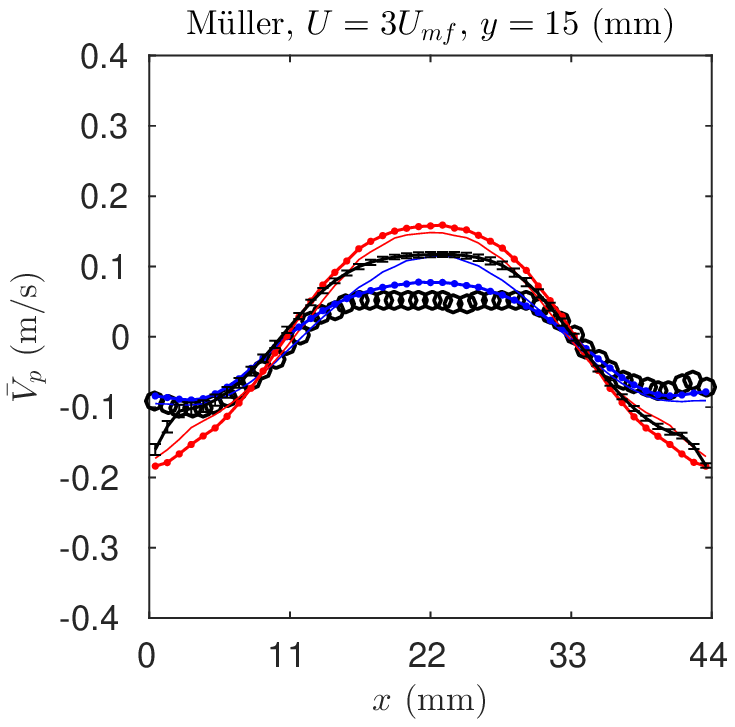}
\centering\includegraphics[height=0.3\linewidth]{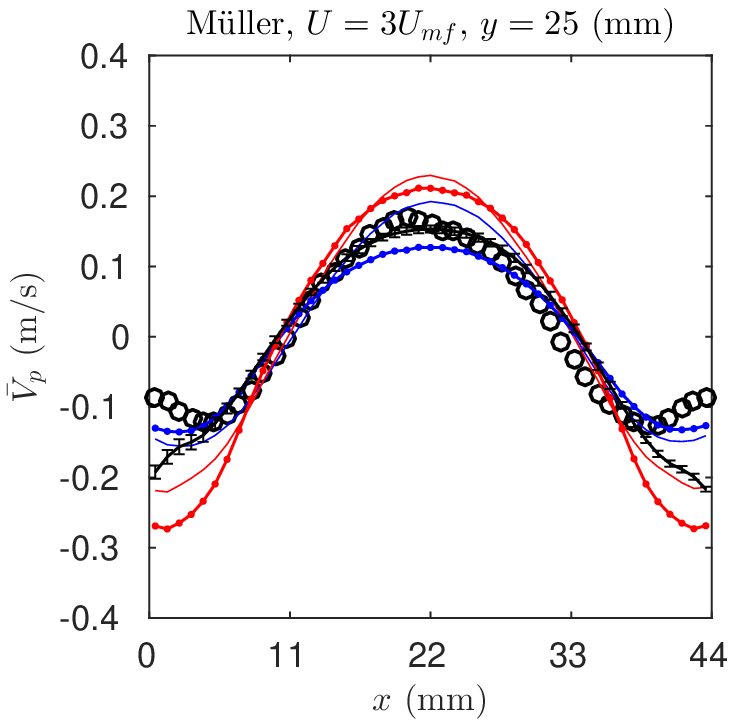}
\centering\includegraphics[height=0.3\linewidth]{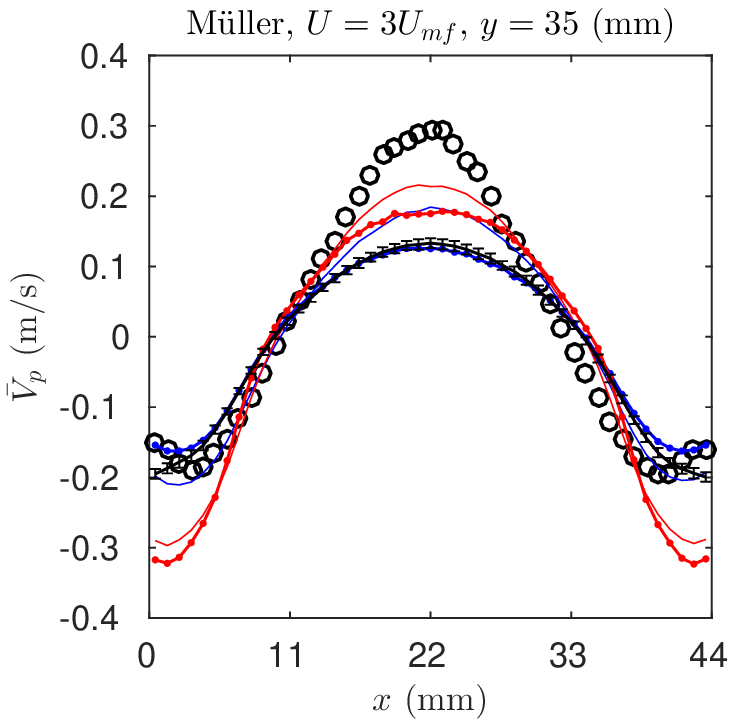}
\caption{(Color online.) Time-averaged mean void fraction (top row) and streamwise particle velocity (bottom row) profiles for the M{\"u}ller bed.}
\label{fig.mueller}
\end{figure}

Figure~\ref{fig.mueller} gives the comparison of the different models and experimental data for the M{\"u}ller bed. The (gas) void fraction profiles are quite similar to previous results benchmarking against this dataset. At the lower $y$ = 16.4~mm elevation, the slug profile is quite flat which the CFD-DEM models are able to reproduce. However, at the higher elevation of $y$ = 31.2~mm the profile shows more variation and the models are unable to capture the high solids concentration (low void fraction) along the walls. This result is consistent with both the original simulation results of M{\"u}ller et al. \cite{mueller09} and the original MFiX-DEM validation results \cite{li12}. We note that the solids concentration observed in the data near the walls (specifically the left wall at $y$ = 31.2~mm) approaches the maximum random packing limit of mono-dispersed, spherical particles.

The velocity profiles in Fig.~\ref{fig.mueller} are also quite similar to previously published results \cite{mueller08, li12}. The most basic trends are captured well by the MFIX-classic and MFIX-Exa models: particles move up in the center with the slugs and fall back down along the walls with the profiles becoming gradually more sharp with increasing elevation from the inlet. The CFD-DEM models have a tendency to over-predict the centerline velocity at $y$ = 15~mm, agree at $y$ = 25~mm, and then under-predict at the $y$ = 35~mm elevation.  A lower, broader $V_p$ at $y$ = 35~mm was also observed in recent particle-resolved DNS of the M{\"u}ller bed \cite{luo16}. However, PR-DNS did a better job of predicting the flat center of the $y$ = 0.15~mm velocity profile than the CFD-DEM models. The present results fail to capture the sharp up-turn near the walls, again, similar to previous CFD-DEM results of this case \cite{mueller08,li12}.

%%
%%  link
%%
\subsection{Link spout-fluidized bed}
\label{sec.link}
The third benchmark case is the spout-fluidized bed of Link et al.  \cite{link08}, hereafter referred to as the Link bed. Unlike the previous two cases, the Link bed is not ``pseudo-2D,'' with a bed depth (84~mm) over half the width (154~mm). Further, the Link bed is the only benchmark case studied here with a nonuniform gas inlet. Instead, a 22~mm wide, 12~mm deep high velocity spout region is centered on the inlet plane, which is surrounded by a lower velocity gas distributor for uniform fluidization. The dual-inlet bed allows the sweeping of a 2-D flow-regime map with spouted bed and fluidized bed behavior as its axes \cite{link08}. Experimental data from three conditions in the flow regime map were provided: 
\begin{itemize}
    \item case B1: $U_{in} = 2.5$~m/s, $U_{spout} = 60$~m/s, flow regime: intermediate spout-fluidization, 
    \item case B1: $U_{in} = 2.5$~m/s, $U_{spout} = 90$~m/s, flow regime: spouting with aeration, 
    \item case B1: $U_{in} = 3.5$~m/s, $U_{spout} = 65$~m/s, flow regime: jet in fluidized bed. 
\end{itemize}
Also unique to the Link bed is the measurement technique which used positron emission particle tracking (PEPT) to collect time-averaged mean, $\bar{V}_p$, and fluctuating, i.e., standard deviation, $V'_p$, particle velocity profiles. The profiles were collected at two elevations, $y = 15$ and 25 mm. The spatial averaging region is the same depth of the spout inlet, but covers the full width of bed and is assumed to have a vertical range of $\pm 5$~mm.

\begin{table}[!ht]
  \begin{center}
    \caption{Simulation parameters of the Link bed.}
    \label{t.link}
    \begin{tabular}{lll}
    \hline
      \bf{Bed properties} &  &  \\
      Width & $L_x$ & 154 (mm) \\
      Height & $L_y$ & 1000 (mm) \\
      Depth & $L_z$ & 84 (mm) \\
      Grid & $\Delta^*$ & 1.59 \\
      \bf{Particle properties} &  &  \\
      Number & $N_p$ & 44800 \\
      Diameter & $d_{p}$ & 4.04 (mm) \\
      Density & $\rho_{p}$ & 2526 (kg/m$^3$) \\
      \bf{Collision properties} &  &  \\
      Restitution coeff. & $e_{pp}$, $e_{pw}$ & 0.97 \\
      Friction coeff. & $\mu_{pp}$, $\mu_{pw}$ & 0.1 \\
          Spring stiffness & $k_n$ & 43000 (N/m) \\
      \bf{Fluid properties} &  &  \\
      Density & $\rho_g$ & 1.2 (kg/m$^3$) \\
      Viscosity & $\mu_g$ & $1.8 \times 10^{-5}$ (Pa-s) \\
      \hline
    \end{tabular}
  \end{center}
\end{table}

As in the M{\"u}ller bed, particle properties needed for CFD-DEM simulations were measured and reported, see Table~\ref{t.link}. The CFD grid in this case is slightly finer than the  $\Delta^* \approx 2$ guideline which was required to resolve the spout with an even number of cells, here $2 \times 2$. The time scale of bed hydrodynamics are considerably faster in the Link bed than the other uniformly fluidized beds. Therefore, 2s bins are sufficient for time-averaging the velocity profiles. Again 12 non-overlapping bins are used to to collect statistics beginning after a 2s start-up period. No ramping of the inlet velocity is used for simulations of the Link bed.

\begin{figure}[htb]
\centering\includegraphics[height=0.3\linewidth]{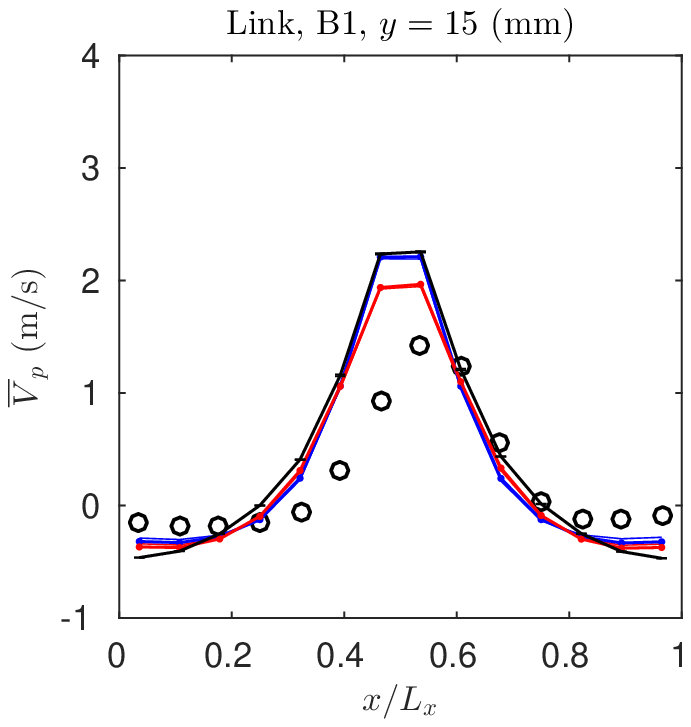}
\centering\includegraphics[height=0.3\linewidth]{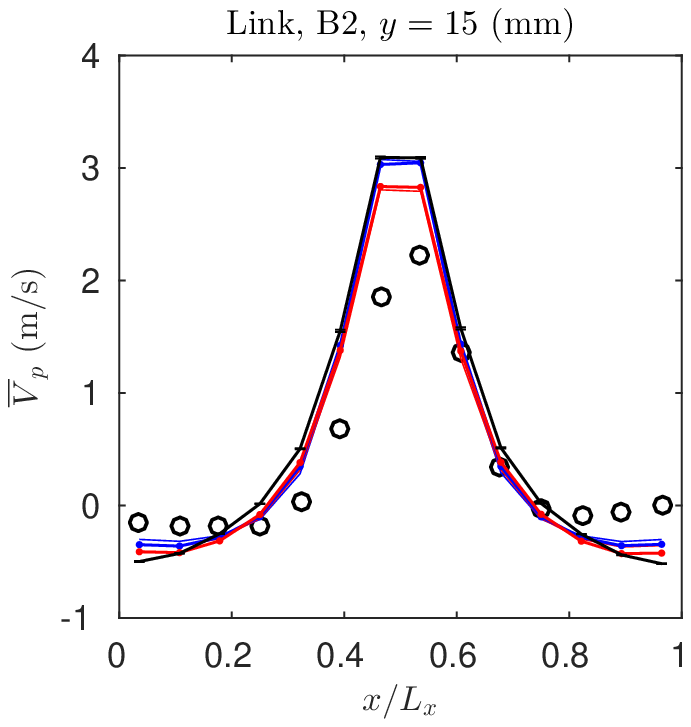}
\centering\includegraphics[height=0.3\linewidth]{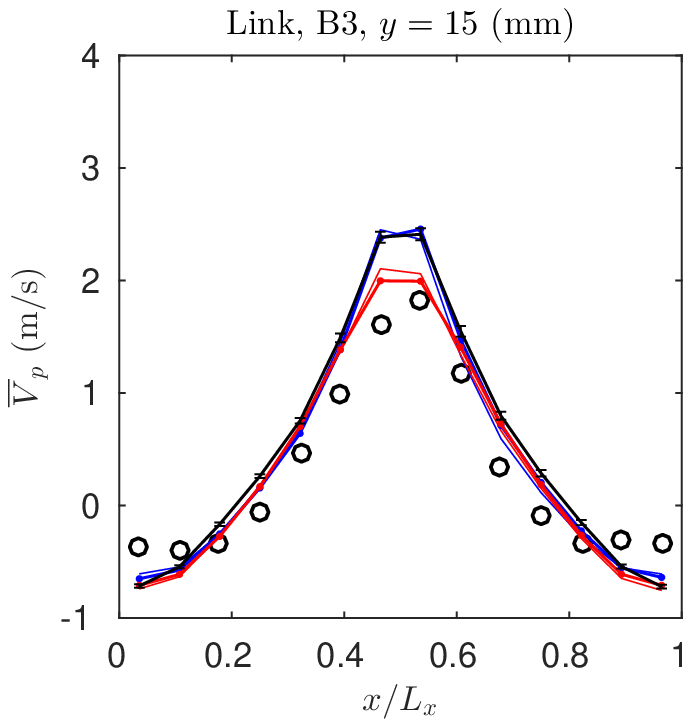} \\
\centering\includegraphics[height=0.3\linewidth]{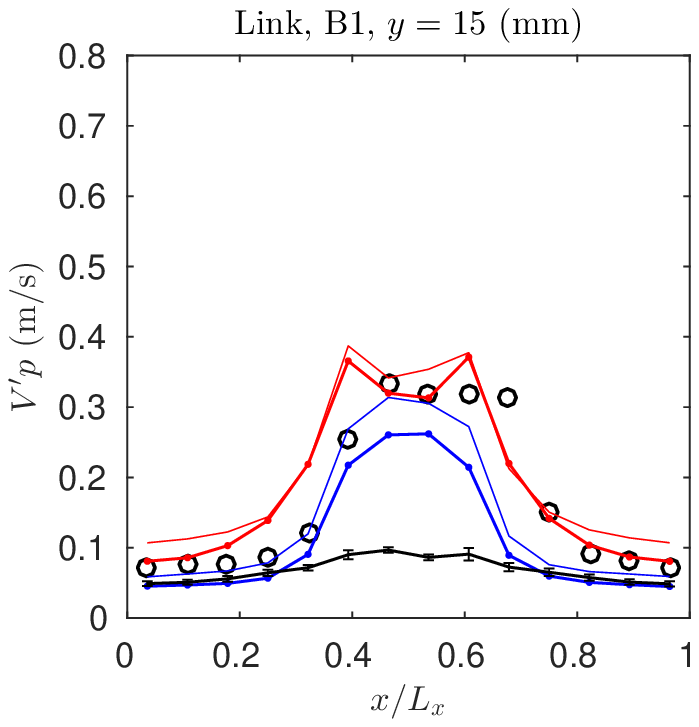}
\centering\includegraphics[height=0.3\linewidth]{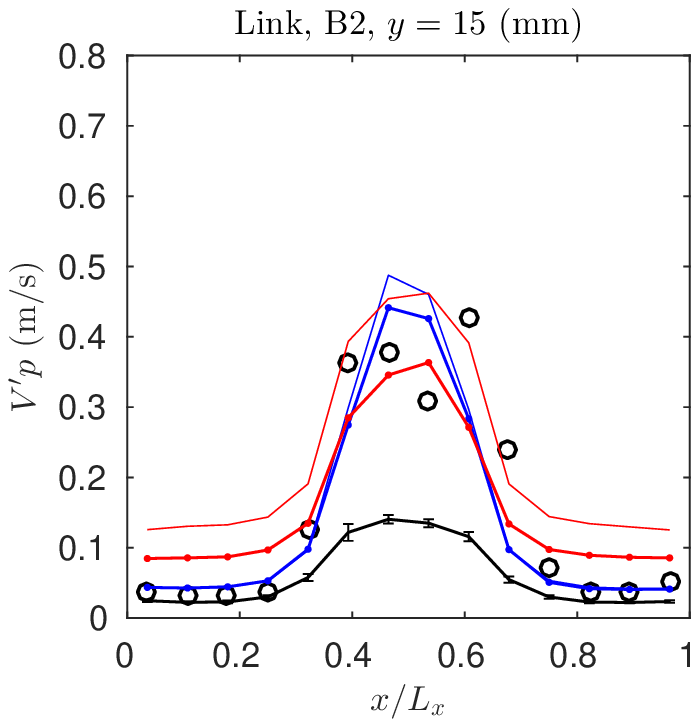}
\centering\includegraphics[height=0.3\linewidth]{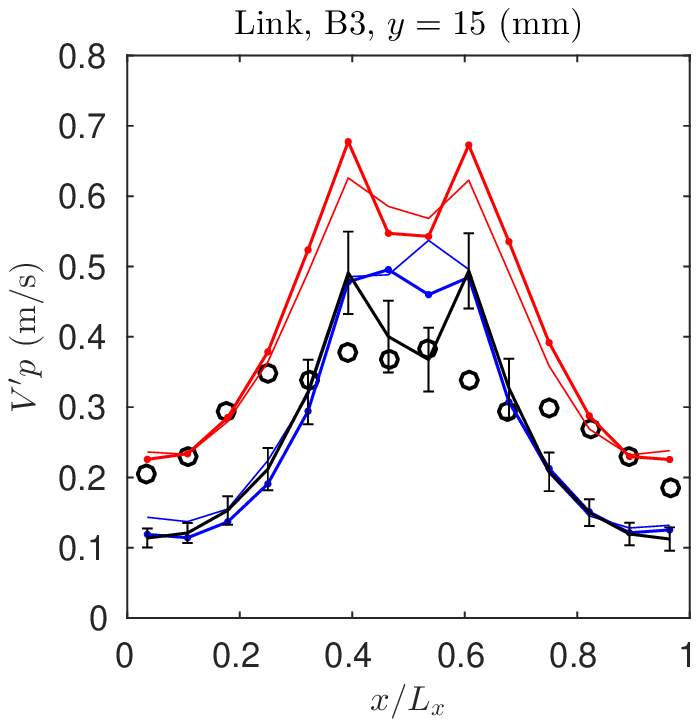}
\caption{(Color online. See Fig.~\ref{fig.mueller} for key.) Time-averaged mean (top row) and fluctuating (bottom row) streamwise particle velocity profiles at the lower elevation in the Link bed.} 
\label{fig.link.y1}
\end{figure}

\begin{figure}[htb]
\centering\includegraphics[height=0.3\linewidth]{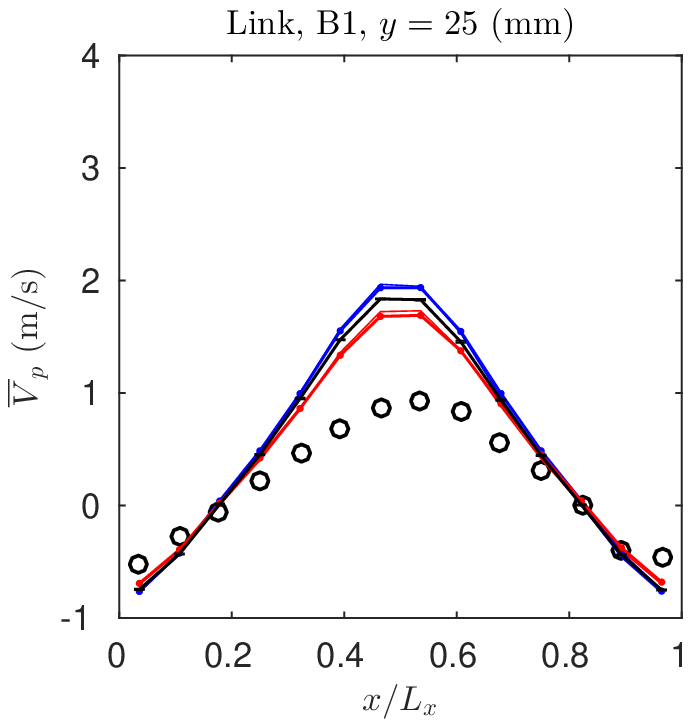}
\centering\includegraphics[height=0.3\linewidth]{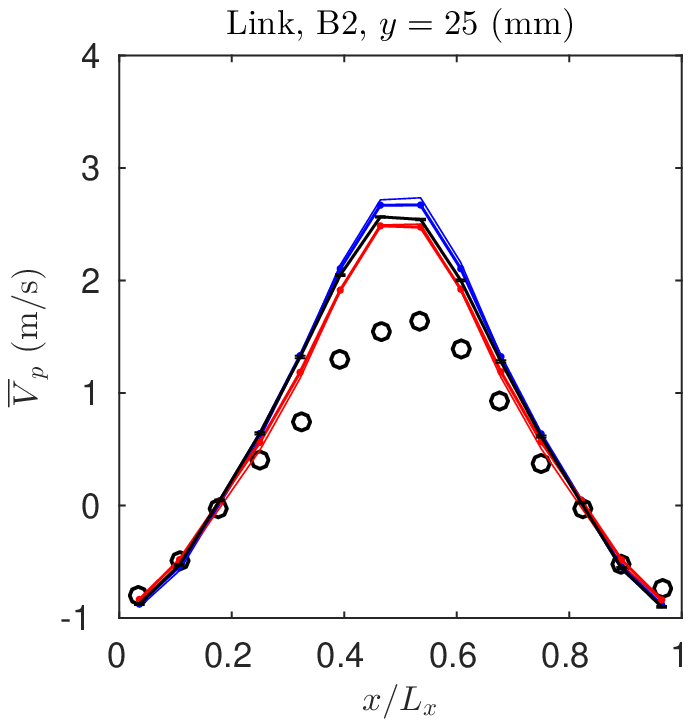}
\centering\includegraphics[height=0.3\linewidth]{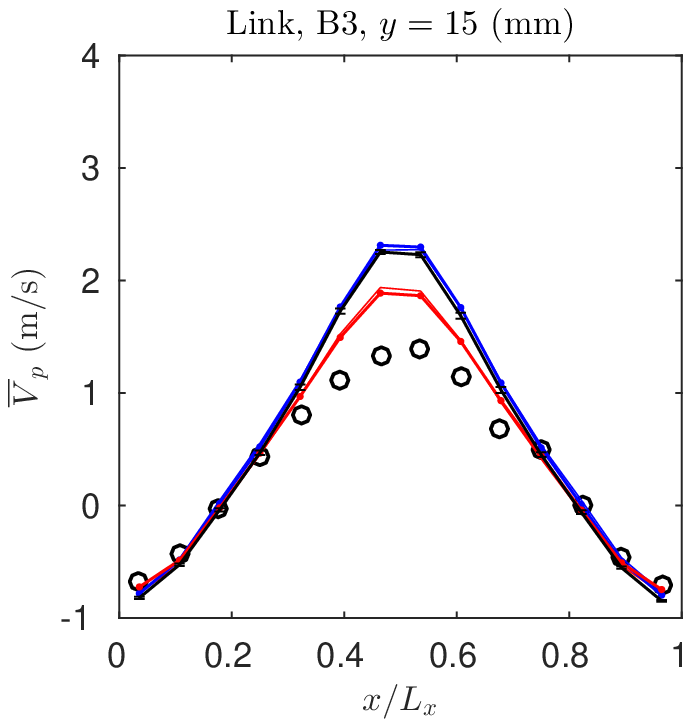} \\
\centering\includegraphics[height=0.3\linewidth]{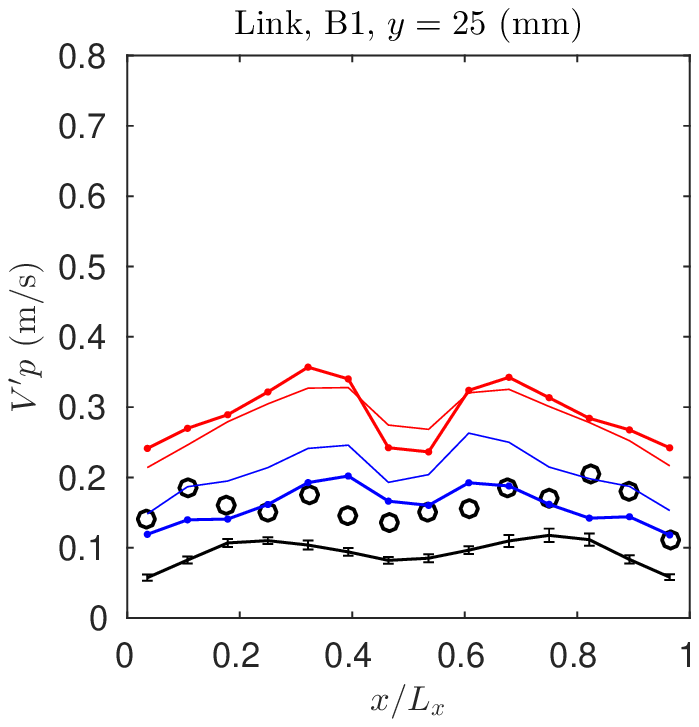}
\centering\includegraphics[height=0.3\linewidth]{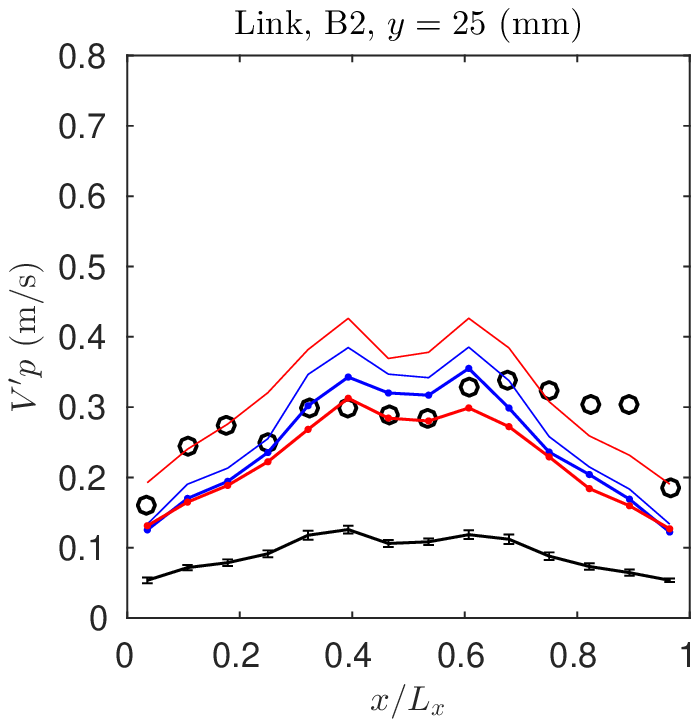}
\centering\includegraphics[height=0.3\linewidth]{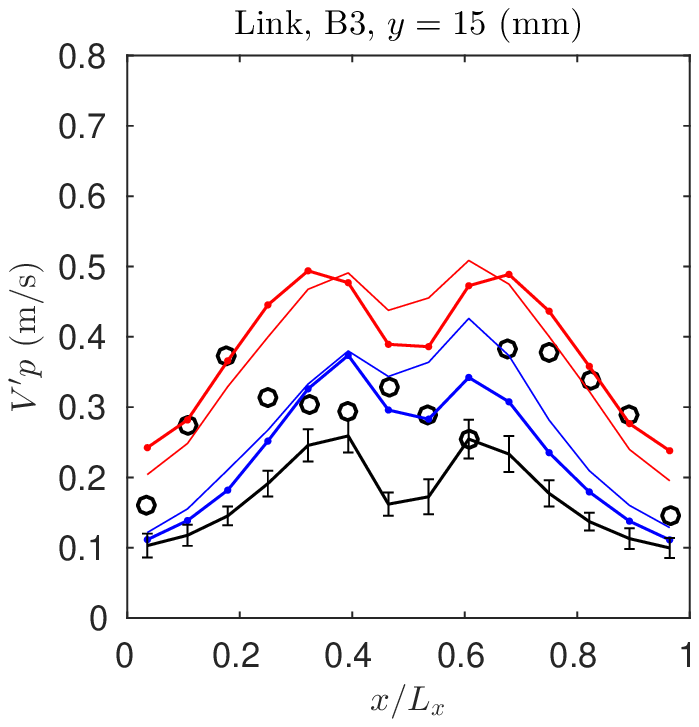}
\caption{(Color online. See Fig.~\ref{fig.mueller} for key.) Time-averaged mean (top row) and fluctuating (bottom row) streamwise particle velocity profiles at the upper elevation in the Link bed.} 
\label{fig.link.y2}
\end{figure}

The velocity profiles for all cases at the lower and upper elevations are shown in Figs.~\ref{fig.link.y1} and \ref{fig.link.y2}, respectively. Mean velocity profiles compare favorably with little spread among the numerical solutions. The mean velocity profiles for cases B1 and B2 were also reported in the original MFiX CFD-DEM validation study \cite{li12}. Compared to previous results \cite{link08, li12}, the present solutions over-predict the mean particle velocity in the central jet region at the upper elevations. This could be due to the increased CFD resolution considered here. The fluctuating particle velocity profiles show more spread among the numerical solutions, which perhaps is to be expected. Generally, the classic MFiX results do a good job predicting this measurement. However, the preliminary MFiX-Exa code shows several discrepancies, particularly for condition B2 and at the 15~mm elevation of condition B1. Again, we are not so concerned that the numerical solution deviates from the experimental data, but here it also deviates from all four MFiX 2016.1 simulations performed as well as previous results. It is believed that this discrepancy is due to the simplified tangential collision model because the difference is largest in the most spouted regions, i.e., condition B2 and the lower elevation, and previous works have shown the sensitivity of spouted bed simulation results to the collision model \cite{goniva12}. %{\wdf we should test this, let's plan to submit first.}

%%
%% sscp
%%
\subsection{SSCP-I fluidized bed}
\label{sec.sscp}
The last benchmark case considered for the preliminary MFiX-Exa code is the first small-scale challenge problem carried out at the National Energy Technology Laboratory \cite{gopalan16}. Referred to as the SSCP-I bed, the goal of the challenge problem was to collect high fidelity experimental data including uncertainty with all material, flow and geometrical parameters required for numerical modeling measured and reported. The geometrical and material properties of the SSCP-I bed are given in Table~\ref{t.sscp}. Two different types of experimental data were collected. First, the bed pressure drop, $DP_{bed} = p_g(y_1) - p_g(y_2)$, was measured between elevations of $y_1 = 41.3$~mm and $y_2 = 346.1$~mm; both time-averaged mean and standard deviations were reported. Second, the bed dynamics were recorded with high speed video which was analyzed using particle tracking velocimetry (PTV). The PTV calculated particle velocities are binned into five spatial regions spanning the width of the bed, each bin with a square edge length of 45.7~mm centered at a height of 76.2~mm. Two types of spatial statistics were collected, reported as ``Eulerian'' and ``Lagrangian'' statistics. Because both methods provide similar results, we choose to use the ``Eulerian'' statistics which are similar to the spatial bin averaging used in the other benchmark studies in this work
%, i.e., particle velocities are first averaged in their spatial region {\jmhm \st{or box-averaged}} and then time-averaged for mean and fluctuating data. 
Bin-averaged vertical and horizontal velocities were reported as mean and fluctuating statistics.

\begin{table}[!ht]
  \begin{center}
    \caption{Simulation parameters of the SSCP-I bed.}
    \label{t.sscp}
    \begin{tabular}{lll}
    \hline
      \bf{Bed properties} &  &  \\
      Width & $L_x$ & 230 (mm) \\
      Height & $L_y$ & 1220 (mm) \\
      Depth & $L_z$ & 75 (mm) \\
      Grid & $\Delta^*$ & 1.94 \\
      \bf{Particle properties} &  &  \\
      Number & $N_p$ & 92948 \\
      Diameter & $d_{p}$ & 3.256 (mm) \\
      Density & $\rho_{p}$ & 1131 (kg/m$^3$) \\
      \bf{Collision properties} &  &  \\
      Restitution coeff. & $e_{pp}$, $e_{pw}$ & 0.84, 0.92 \\
      Friction coeff. & $\mu_{pp}$, $\mu_{pw}$ & 0.35 \\
      Spring stiffness & $k_n$ & 1000 (N/m) \\
      \bf{Fluid properties} &  &  \\
      Density & $\rho_g$ & 1.2 (kg/m$^3$) \\
      Viscosity & $\mu_g$ & $1.8 \times 10^{-5}$ (Pa-s) \\
      \hline
    \end{tabular}
  \end{center}
\end{table}

No-slip walls are applied to all four vertical walls because  $\Delta^* \approx 2$ allows for a cross-sectional gird of 36x12 CFD cells. The uniform inlet velocity is linearly ramped over one second from $U_{mf}$ to $U_{in}$. Three conditions are considered corresponding to $U_{in} = 2$, 3 and $4 U_{mf}$, where $U_{mf}$ was measured as 1.095~m/s. Similar to the Goldschmidt and M\"{u}ller beds, twelve 5s time-averaging bins are used to collect statistics starting after a 5s transient period. The bed pressure drop is calculated by averaging gas-phase pressure over the two planes of CFD cells above and below the desired locations and then linearly interpolating to the $y_1$ and $y_2$ locations. As in the Goldschmidt bed, the five spatial averaging regions only consider particles with $z_i \le 1.5 d_p$. The approximation of a $1.5 d_p$ depth of view was both reported \cite{gopalan16} and verified by comparing particle counts with unpublished data.

\begin{figure}[htb]
\centering\includegraphics[width=0.5\linewidth]{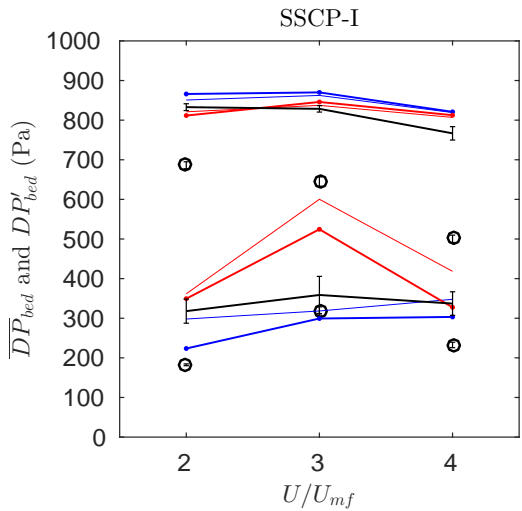}
\caption{(Color online. See Fig.~\ref{fig.mueller} for key.) Time-averaged mean and fluctuating bed pressure drop in the SSCP-I bed.} 
\label{fig.sscp.dp}
\end{figure}

The SSCP-I bed pressure drop measurements and numerical results are reported in Fig.~\ref{fig.sscp.dp} for the three conditions. The preliminary MFiX-Exa code, consistent with the other simulations, over-estimates the mean bed pressure drop. It is worth noting that the mean pressure drop predicted by the simulations is consistent with the weight of the bed being fully fluidized and that the experimentally measured values are well below this number. 
%{\jmhm \emph{(Maybe give the percentage for one of the cases.)} 
This suggests that the bed may not have been fully fluidized in the experiments, i.e., a portion of the bed weight may have been mechanically supported. 
%\emph{(We don't have sufficient, documented data to say much more than this.)}} 
The fluctuating bed pressure drop acceptably predicts the experimental data and agrees with other numerical results, however the classic MFiX results show considerable spread, most severely at $3U_{mf}$. The results in Fig.~\ref{fig.sscp.dp} suggest that perhaps the large CFD timestep allowed by the ``implicitly'' coupled MFiX classic models has led to considerable numerical error. To test this hypothesis, both ``implicit'' cases (i.e. MFiX Release 2016.1 with the \texttt{GARG\_2012} kernel and FOU and SMART variable extrapolation) were re-run using the same $dt_{CFD} \le \tau_{coll}$ restriction of the explicit models. Reducing the maximum allowable timestep caused the predicted $DP'_{bed}$ in these cases to decrease from 525 $\pm$ 62~Pa to 323 $\pm$ 52~Pa for FOU and from 600 $\pm$ 25~Pa to 489 $\pm$ 66~Pa. 
%% garg smart: 
%% 837.37    10.86    6.003293e+02    2.503552e+01
%% garg smart limit dt:
%% 852.673    11.20    4.890314e+02    6.620490e+01
%%
%% garg fou: 
%% 8.459468e+02    1.307261e+01    5.245200e+02    6.187234e+01
%% garg fou limitDT: 
%% 8.655310e+02    4.738216e+00    3.226142e+02    5.202764e+01

\begin{figure}[htb]
\centering\includegraphics[height=0.3\linewidth]{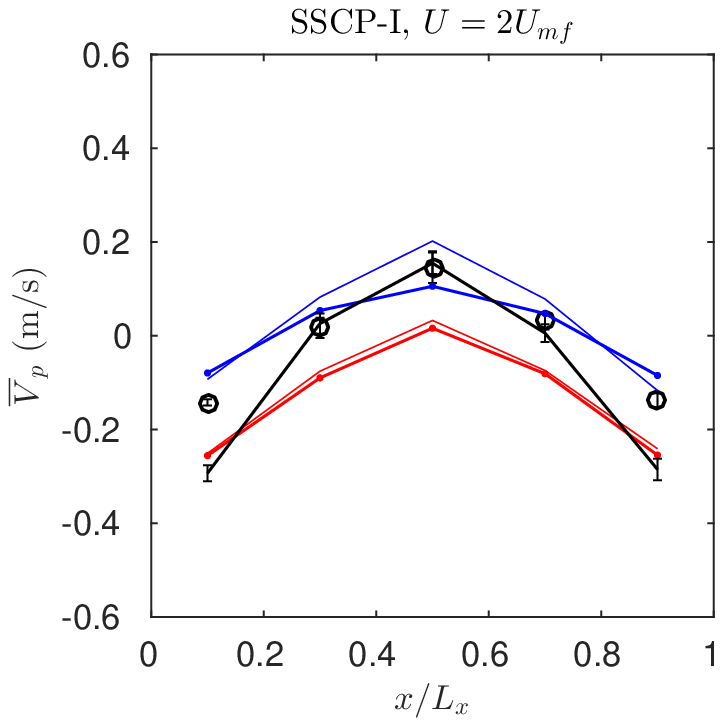}
\centering\includegraphics[height=0.3\linewidth]{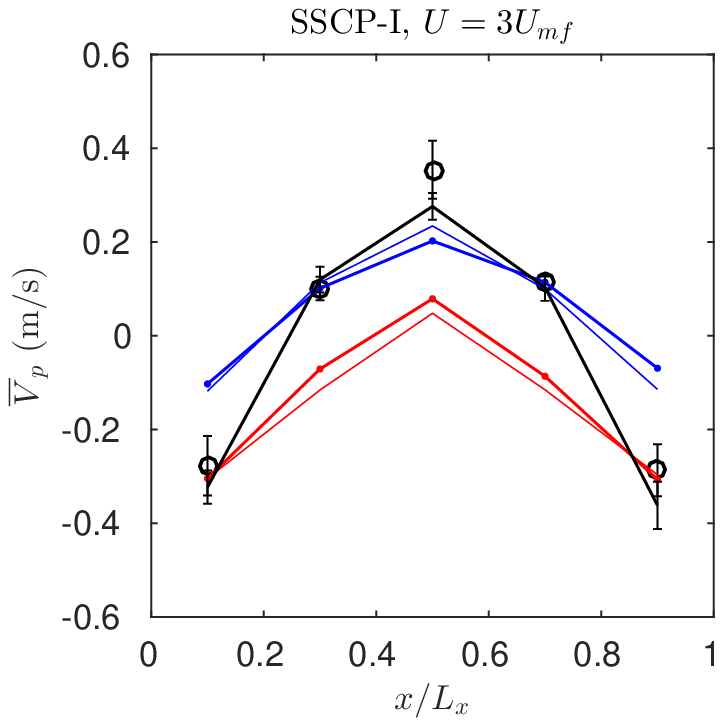}
\centering\includegraphics[height=0.3\linewidth]{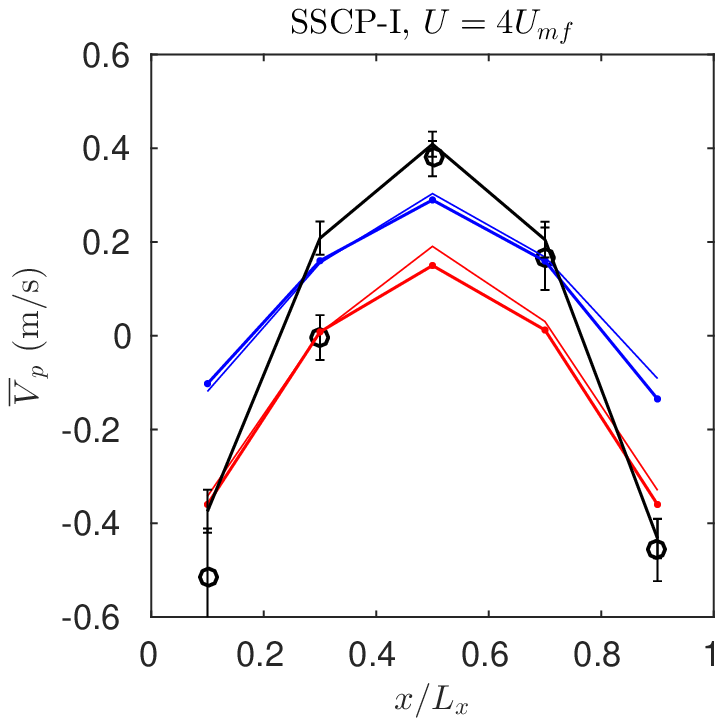} \\
\centering\includegraphics[height=0.3\linewidth]{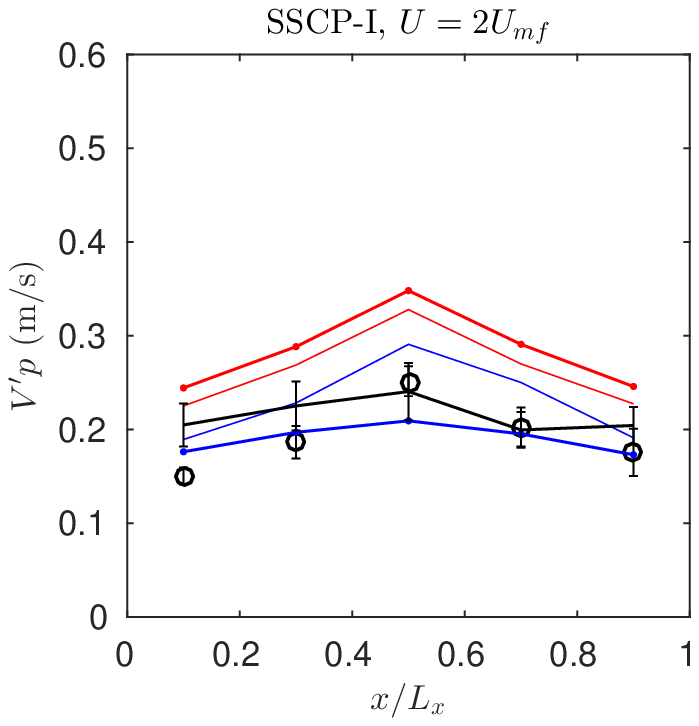}
\centering\includegraphics[height=0.3\linewidth]{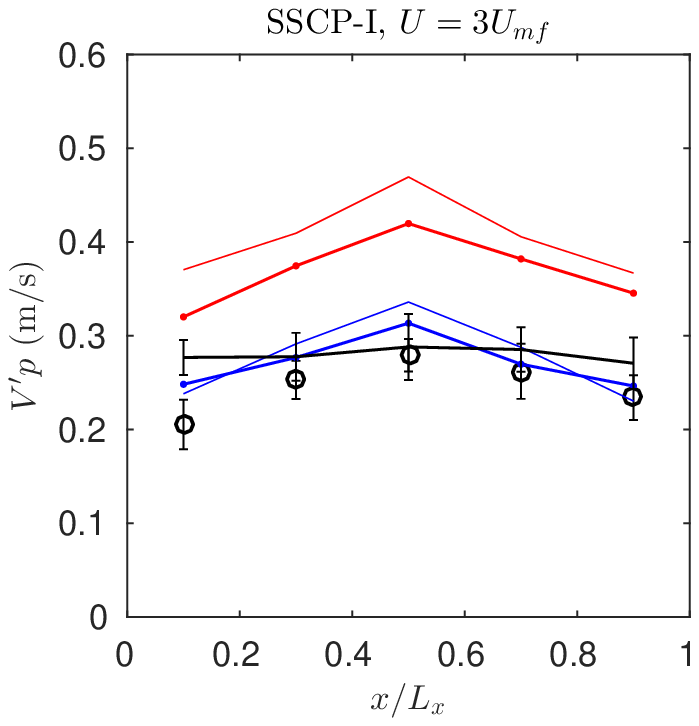}
\centering\includegraphics[height=0.3\linewidth]{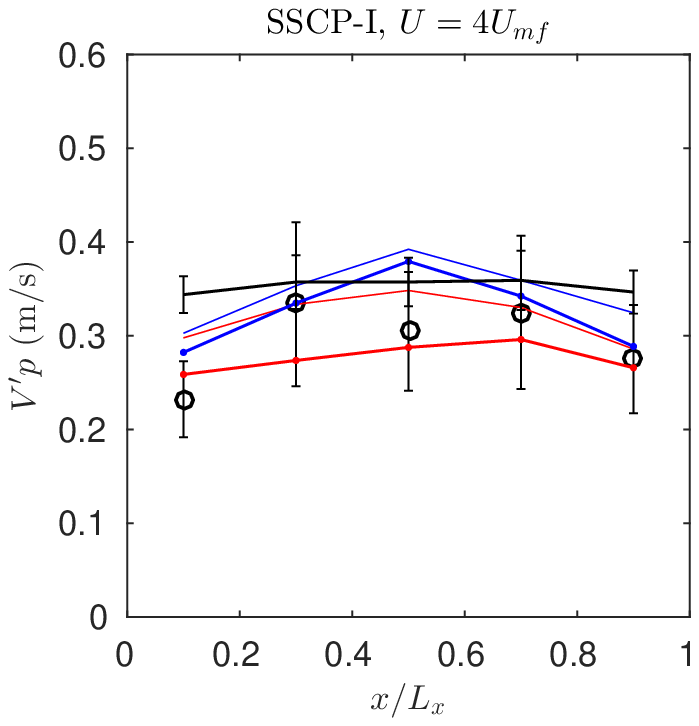}
\caption{(Color online. See Fig.~\ref{fig.mueller} for key.) Time-averaged mean (top row) and fluctuating (bottom row) streamwise particle velocity profiles in the SSCP-I bed.}
\label{fig.sscp.v}
\end{figure}

\begin{figure}[htb]
\centering\includegraphics[height=0.3\linewidth]{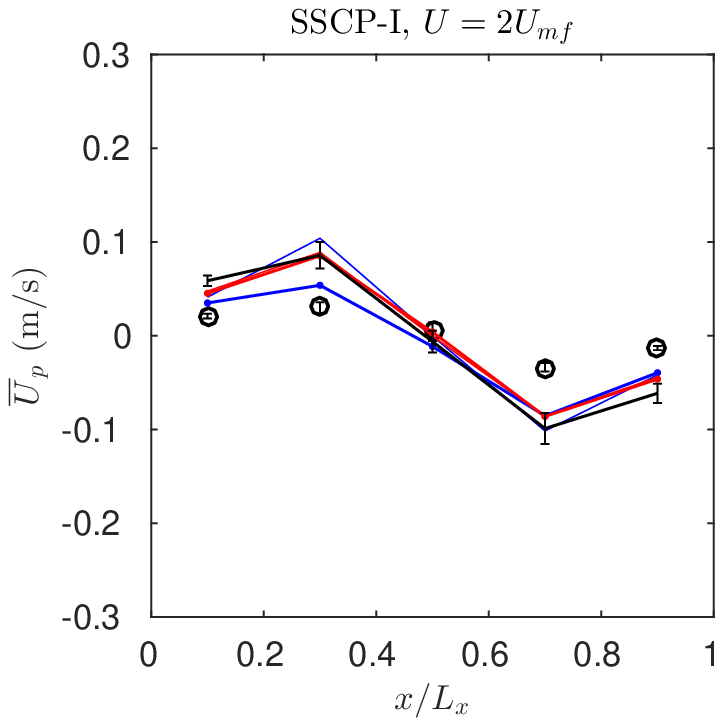}
\centering\includegraphics[height=0.3\linewidth]{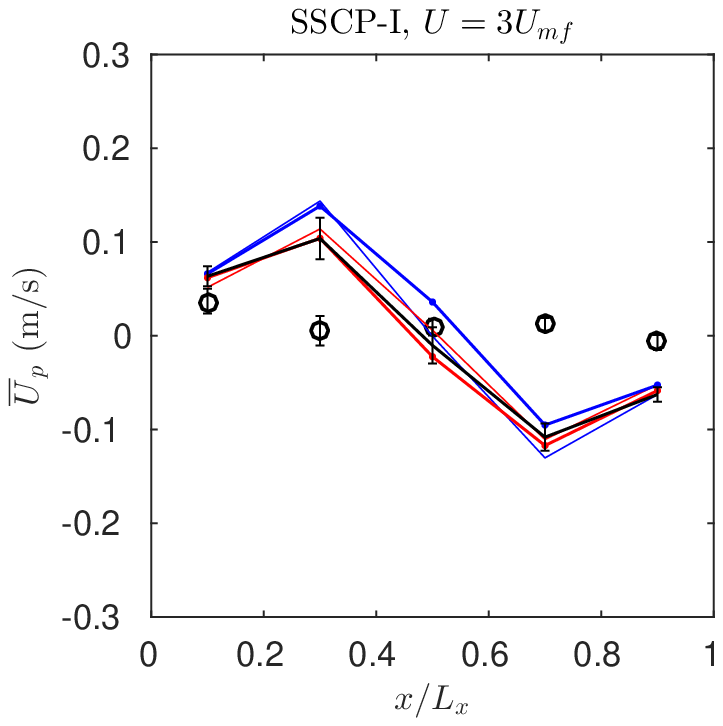}
\centering\includegraphics[height=0.3\linewidth]{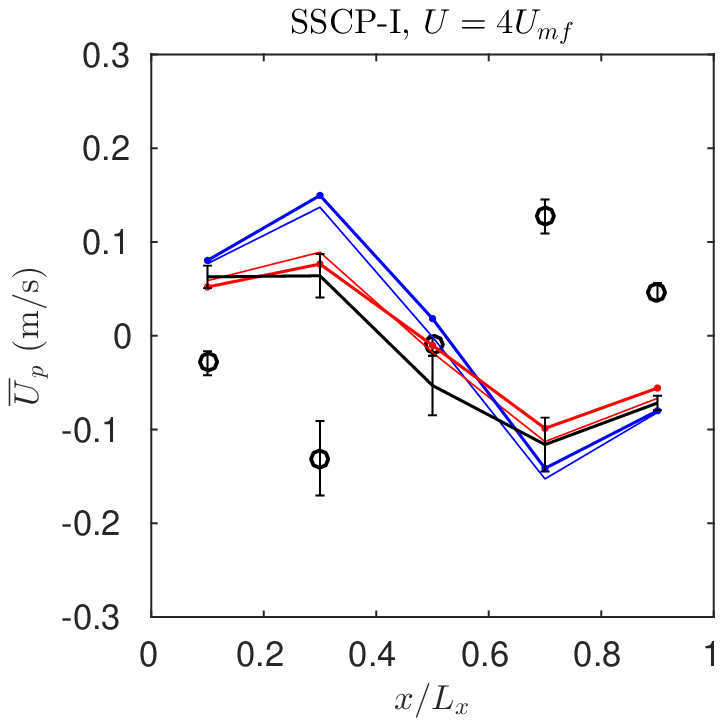} \\
\centering\includegraphics[height=0.3\linewidth]{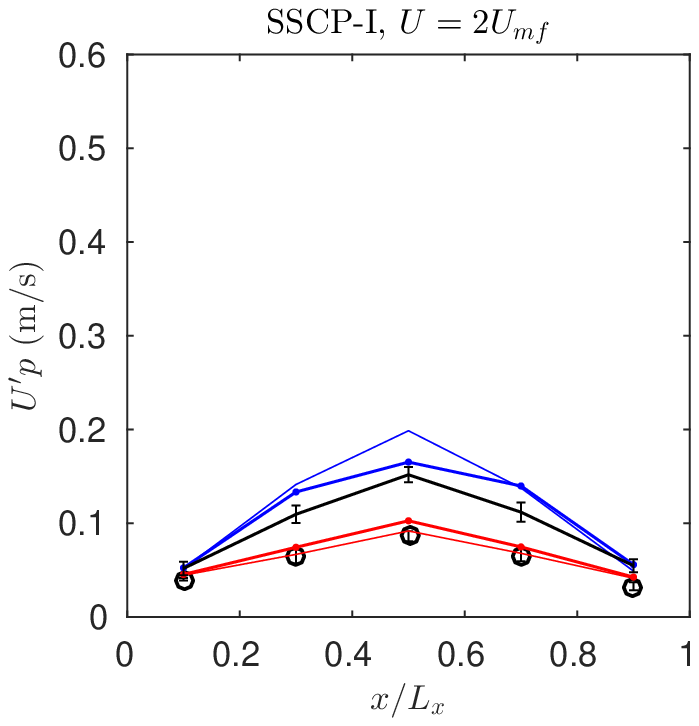}
\centering\includegraphics[height=0.3\linewidth]{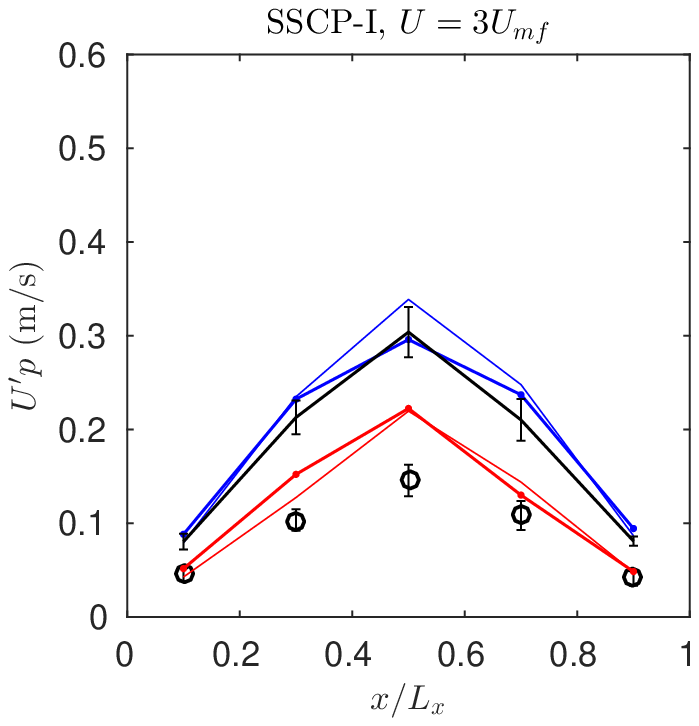}
\centering\includegraphics[height=0.3\linewidth]{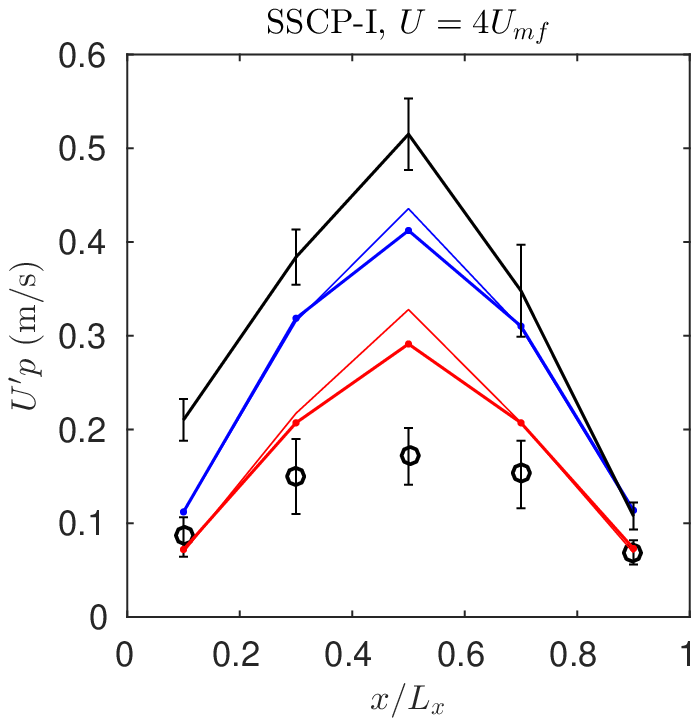}
\caption{(Color online. See Fig.~\ref{fig.mueller} for key.) Time-averaged mean (top row) and fluctuating (bottom row) transverse particle velocity profiles in the SSCP-I bed.}
\label{fig.sscp.u}
\end{figure}

The vertical and horizontal velocity profiles are compared in Figs.~\ref{fig.sscp.v} and \ref{fig.sscp.u}, respectively. Generally, the preliminary MFiX-Exa code compares favorably to the experimental data and shows reasonable agreement with the other numerical solutions. The most noticeable discrepancy in particle velocity is in the mean transverse velocity. As $U_{in}$ increases from 2 to $4 U_{mf}$, the measured $\bar{U}_p$ diverges from all presented numerical results. Similar observations have been reported previously \cite{elghannay14, koralkar16, liu19}, again making the seemingly bad validation result a success in terms of code-to-code benchmarking the preliminary MFiX-Exa code. We note that the disagreement with experiment is not quite as poor as indicated in Fig.~\ref{fig.sscp.u}. The physical location of the imaging window clips a portion of counter-rotating vortices, the center of which is predicted lower than measured experimentally.

%%
%%  reproducible
%%
\subsection{Reproducibility}
\label{sec.reproducible}
\begin{figure}[ht]
\centering\includegraphics[width=0.5\linewidth]{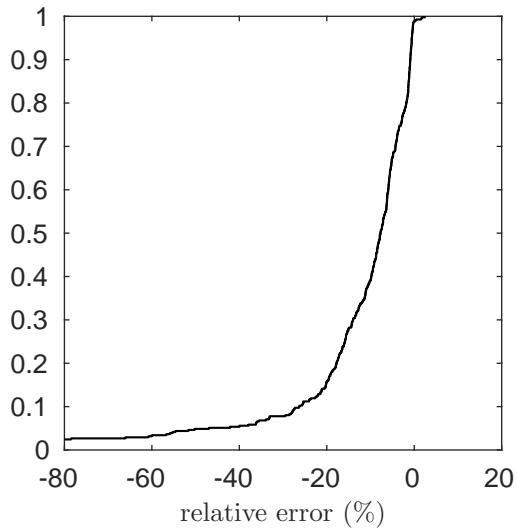}
\caption{Reproducibility of the MFiX-Exa benchmark simulations shown as a cumulative distribution function of relative errors (negative error indicates statistically similar results).}
\label{fig.reprod}
\end{figure}

The MFiX and MFiX-Exa codes and development histories are archived at \url{http://mfix.netl.doe.gov/gitlab}. Results presented in this paper were generated with the MFIX \texttt{Release-2016-1} branch and MFIX-Exa \texttt{18.10} tagged version. All necessary source code modifications, input decks and post-processing scripts are collected in a separate gitlab repository specific for this work. Originally, MFiX-Exa simulations were run on several developmental branches. To test the reproducibility of the results, all MFiX-Exa simulations were repeated on the tagged \texttt{18.10} code, the results of which are presented in this work. A comparison of the original and repeated results is agglomerated in Fig.~\ref{fig.reprod} in which the relative error including statistical uncertainty, 
\beq
\textrm{error} = \frac{{\left| a - b \right| - \delta_a - \delta_b }}{{\left(a + b \right)/2}} , 
\label{eq.aberr}
\eeq
of each data point presented in this work. Negative relative errors indicate overlapping errorbars, i.e., statistically similar results, which occurs in a vast majority of cases, in fact, more often than expected for the 95\% confidence level used in the errorbars.

%%
%%  summary
%%
\section{Summary}
\label{sec.summary}
This work presents an initial benchmarking study of the preliminary MFiX-Exa code, i.e., a refactoring of the cold-flow MFiX-DEM (\url{mfix.netl.doe.gov}) code into the AMReX (\url{amrex-codes.github.io}) framework. The preliminary code will serve as a starting point for extensive ongoing and future development of the code into an exascale-capable application under the US DOE's Exascale Computing Project (\url{www.exascaleproject.org}). Each case is simulated with the preliminary MFiX-Exa code as well as four classic MFiX-DEM models using ``implicit'' \texttt{GARG\_2012} kernel and explicit \texttt{SQUARE\_DPVM} transfer kernels with first-order upwind and MUSCL variable extrapolation.

Four benchmark problems were considered, all of which correspond to previously published physical experiments: the Goldschmidt \cite{goldschmidt04} and M{\"u}ller \cite{mueller08, mueller09} fluidized beds, the Link \cite{link08} spout-fluid bed, and NETL's SSCP-I \cite{gopalan16} fluidized bed. By and  large, the preliminary MFiX-Exa code compares favorably to the classic MFiX-DEM results and reasonably well to the experimental data with only two noticable outliers. For the Goldschmidt bed at the lowest superficial velocity, $U_{in} = 1.25 U_{mf}$, the fluctuating bed height is roughly twice that of the next largest prediction. It was determined that this over-prediction is due to the system locking into a nearly periodic (i.e., not chaotic) slugging pattern. It is believed that this behavior (atypical for many particle systems) is related to the thinness of the bed and the low-order numerical methods, which are almost exclusively formally first-order in the preliminary MFiX-Exa code. The regular pattern may only be expected to occur in a narrow operating regime close to minimum fluidization and therefore is not considered a significant discrepancy. A second noticable code-to-code difference occurs in the lower jet region of the Link spout-fluid bed. This discrepancy is believed to originate from the simplified tangential force model in the LSD collision model. As discussed in Sec.~\ref{sec.model.lsd}, the improved computational efficiency of the simplified model does come with a slight decrease in model accuracy.

%%
%%  todo
%%
\section{Future Outlook}
\label{sec.todo}
Significant development has already taken place on the MFiX-Exa code. Perhaps most significantly, the SIMPLE algorithm has been replaced with a cell-centered, low-Mach Number projection method and an improved Embedded Boundary method has been implemented in AMReX, allowing MFiX-Exa to consider non-rectangular geometries. This substantial code overhaul will be reported on in the near future.  Active development includes load balancing strategies to improve parallel performance and  geometry-dependent adaptive mesh refinement for improved wall resolution. Beyond repeating the relatively simple cases considered in this work, future MFiX-Exa V\&V work will include extending the benchmarking database to more complex cases, for example considering
\begin{itemize}
    \item integer-disperse material, e.g., \citet{goldschmidt03, jiang18} 
    \item simple, non-rectangular geometries, e.g., \citet{boyce16, penn17} 
    \item complex geometries, e.g., \citet{xu18, fullmer18a, jalali18}
    \item ordered pattern formation, e.g., \citet{wu16, bakshi18}
\end{itemize}

\section*{Acknowledgment}
This research was supported by the Exascale Computing Project (17-SC-20-SC), a collaborative effort of the U.S. Department of Energy Office of Science and the National Nuclear Security Administration.

This work was performed in support of the US Department of Energy’s Fossil Energy Crosscutting Technology Research Program. The Research was executed through the NETL Research and Innovation Center’s MFiX-Exa Support. Research performed by Leidos Research Support Team staff was conducted under the RSS contract 89243318CFE000003.

This work was funded by the Department of Energy, National Energy Technology Laboratory, an agency of the United States Government, through a support contract with Leidos Research Support Team (LRST). Neither the United States Government nor any agency thereof, nor any of their employees, nor LRST, nor any of their employees, makes any warranty, expressed or implied, or assumes any legal liability or responsibility for the accuracy, completeness, or usefulness of any information, apparatus, product, or process disclosed, or represents that its use would not infringe privately owned rights. Reference herein to any specific commercial product, process, or service by trade name, trademark, manufacturer, or otherwise, does not necessarily constitute or imply its endorsement, recommendation, or favoring by the United States Government or any agency thereof. The views and opinions of authors expressed herein do not necessarily state or reflect those of the United States Government or any agency thereof.

\section{References}
\bibliographystyle{model1-num-names}
\bibliography{wdf,addin}

\end{document}